\documentclass[twocolumn]{revtex4} 
\usepackage{amsmath}
\usepackage{amssymb}
\usepackage{graphicx,color}

\begin{document}

\title{
Quantum Singwi-Tosi-Land-Sj\"olander approach for interacting inhomogeneous systems under electromagnetic fields: Comparison with exact results
}

\author{Taichi Kosugi and Yu-ichiro Matsushita}

\affiliation{
Department of Applied Physics, The University of Tokyo, Tokyo 113-8656, Japan
}

\begin{abstract}
For inhomogeneous interacting electronic systems under a time-dependent electromagnetic perturbation,
we derive the linear equation for response functions in a quantum mechanical manner.
It is a natural extension of the original semi-classical Singwi-Tosi-Land-Sj\"olander (STLS) approach for an electron gas.
The factorization ansatz for the two-particle distribution is an indispensable ingredient in the STLS approaches for
determination of the response function and the pair correlation function.
In this study, we choose an analytically solvable interacting two-electron system
as the target for which we examine the validity of the approximation.
It is demonstrated that the STLS response function reproduces well the exact one for low-energy excitations.
The interaction energy contributed from the STLS response function is also discussed.
\end{abstract}


\maketitle

\section{Introduction}

In the 1960s,
Singwi, Tosi, Land, and Sj\"olander (STLS)\cite{bib:3818} proposed a self-consistent scheme to determine the correlation energy for an electron gas
by starting from the classical equation of motion (EOM) for calculating the response to a perturbation potential.
Since the original EOM involves the two-particle distribution of the electron gas,
they introduced an ansatz which assumes that the two-particle distribution can be factorized into
the products of two one-particle distributions and the static pair correlation function $g$ in equilibrium,
in order for the dynamics of weakly perturbed gas to be governed by a linear equation.
They used this approximation and the local-field correction function $G$, which takes into account the effects of electronic interactions ignored in the random phase approximation (RPA),
so that the self-consistent determination of $g$ and the density response function $\chi$ is possible. 
The local-field correction is nothing but the local exchange correlation kernel in the language of the (time-dependent) density functional theory (DFT).
STLS demonstrated numerically that,
despite the violation of the compressibility sum rule,
the considerable improvement of calculated $g$ is achieved compared to the RPA results.
Various local-field correction functions in other forms have been proposed in the literature.\cite{bib:Hubbard_Sham_LFC, bib:3869, bib:3870, bib:3816, bib:3823, bib:3885}
The homogeneous STLS approach was applied also to a spin-polarized electron gas subject to the Zeeman-type coupling.\cite{bib:3828}
Hasegawa and Shimizu\cite{bib:3888, bib:3889} derived the EOM of the Wigner distribution function (WDF)\cite{bib:4117} for an interacting gas
by starting from the Heisenberg equation employing the factorization ansatz as adopted in the original STLS approach.
Their work is thus referred to as a quantum mechanical version of the STLS approach.

The quantitatively accurate predictions and explanations of 
the ground state electronic properties of interacting electronic systems have been successfully realized
by employing DFT with its implementation based on the effective independent-particle equation, called the Kohn-Sham (KS) equation.\cite{bib:77}
TDDFT\cite{bib:3198} has also been formulated to calculate the properties of excited states.
The quality of a practical (TD)DFT calculation depends primarily on the approximation introduced to the exchange correlation functional,
or more specifically the exchange correlation kernel which is generally non-local,
for generating the effective non-interacting system.
It is known, however, that the exchange correlation functionals within local or semi-local approximations proposed so far often fails even qualitatively
to describe long-range correlation effects such as the van der Waals (vdW) interactions.
To overcome the drawbacks in the ordinary DFT framework including such vdW physics,
Dobson et. al.\cite{bib:4075, bib:4076} formulated an inhomogeneous STLS approach for the first time.
They started from the one-particle classical distribution function known as the first Bogoliubov-Born-Kirkwood-Green-Yvon hierarchy equation\cite{bib:4075_Ref_14} to develop their method.
Since they were conscious of the combination of their approach and DFT calculations from the beginning,
it was built so that the quantities directly related to the non-interacting KS system are explicitly involved.
One of the salient features in their approach is the classical vector response function $\boldsymbol{\nu}_0$\cite{bib:4075}
defined to give the non-interacting density response to an applied force.
With that, they were able to derive a Dyson-like integral equation for the interacting response function in which momentum variables have been integrated out.
They demonstrated numerically that their approach for jellium slabs gives the correlation energy in quantitatively good agreement with those obtained by diffusion Monte Carlo calculations.
The applications of their inhomogeneous STLS approach to the metal surface energy\cite{bib:4081}, the metal slabs\cite{bib:4082}, and the spherical atoms\cite{bib:4077}
have also been reported.

Although the applications of semi-classical inhomogeneous STLS approach mentioned above have been reported to give the encouraging results,
there exists necessity to formulate an inhomogeneous STLS approach in a quantum mechanical manner for theoretical consistency.
In the present study,
we therefore propose a quantum STLS approach for inhomogeneous interacting systems under time-dependent electromagnetic perturbation
by deriving the linear equation for response functions in a manner adopted by Hasegawa and Shimizu\cite{bib:3888, bib:3889}.

This paper is organized as follows.
In Section II,
we formulate the quantum STLS approach for inhomogeneous interacting systems under electromagnetic fields,
for which we derive the equation of motion governing the response of one-particle WDF.
In Section III,
we apply the new approach to an interacting two-electron system to capture the behavior of STLS approximation.
We calculate and compare the exact and the approximated results by using the analytically obtained many-body wave functions.

\section{Formalism}

\subsection{Setup}

We consider a three-dimensional interacting electronic system
which is in equilibrium at zero temperature, that is, in the ground state for $t < 0$.
We assume that there is no direct coupling between the spin degree of freedom and magnetic fields.
To examine the response of the system to a time-dependent external electromagnetic field which is turned on at $t = 0$,
we work with the second-quantized Hamiltonian
$
	\hat{H}
	=
		\hat{T}
		+
		\hat{H}_{\mathrm{int}}
		+
		\hat{V}
	,
$
where
\begin{gather}
	\hat{T}
	\equiv
			\sum_\sigma
			\int d^3 r
				\hat{\psi}_{\sigma}^\dagger (\boldsymbol{r}, t)
				T_{\sigma} (\boldsymbol{r}, t)
				\hat{\psi}_{\sigma} (\boldsymbol{r}, t)
	\\
	\hat{H}_{\mathrm{int}}
	\equiv
			\frac{1}{2}
			\sum_{\sigma, \sigma'}
			\int d^3 r d^3 r'
				\hat{\psi}_{\sigma}^\dagger (\boldsymbol{r}, t)
				\hat{\psi}_{\sigma'}^\dagger (\boldsymbol{r}', t)
	\cdot
	\nonumber \\
	\cdot
				v (\boldsymbol{r} - \boldsymbol{r}')
				\hat{\psi}_{\sigma'} (\boldsymbol{r}', t)
				\hat{\psi}_{\sigma} (\boldsymbol{r}, t)
	\label{def_H_int_STLS}
	\\
	\hat{V}
	\equiv
		\sum_\sigma
		\int d^3 r
			\hat{\psi}_{\sigma}^\dagger (\boldsymbol{r}, t)
			V_{\sigma} (\boldsymbol{r}, t)
			\hat{\psi}_{\sigma} (\boldsymbol{r}, t)
	\label{def_V_STLS}
\end{gather}
for the creation $\hat{\psi}_{\sigma}^\dagger$ and the annihilation $\hat{\psi}_{\sigma}$ field operators of an electron having spin $\sigma$.
All the operators in second-quantized form appearing in the present work are in the Heisenberg picture.
The potential applied to the electrons
$V_\sigma (\boldsymbol{r}, t) \equiv V_{0 \sigma} (\boldsymbol{r}) + V_{\mathrm{ext} \sigma} (\boldsymbol{r}, t)$
is the sum of the inherent (present even when $t < 0$) $V_{0 \sigma}$ and the perturbation $V_{\mathrm{ext} \sigma}$ potentials.
$
T_{\sigma} (\boldsymbol{r}, t)
\equiv
[ -i \nabla_{\boldsymbol{r}} + \boldsymbol{A}_{\sigma} (\boldsymbol{r}, t)/c  ]^2/ (2 m)
$
is the gauge covariant kinetic-energy operator for the classical vector potential $\boldsymbol{A}_{\sigma}$,
which we assume is in the Coulomb gauge, that is, its divergence vanishes everywhere. 
$m$ is the electron mass and $c$ is the speed of light.
Although we do not consider spin-dependence of $V_{\sigma}$ and $\boldsymbol{A}_{\sigma}$,
we let them have spin indices
since they make it easy to understand the expressions below when taking the functional derivatives. 
The form of the interaction $v$ between two electrons is not restricted to the ordinary Coulomb type.

\subsection{Equation of motion for Wigner distribution function}

To analyze the quantum dynamics of the interacting system subject to the perturbation,
we introduce the one-particle Wigner distribution operator
\begin{widetext}
\begin{gather}
	\hat{f}_{\sigma} (\boldsymbol{r}, \boldsymbol{p}, t)
	=
		\int
		\frac{d^3 \overline{r}}{ h^3}
			e^{i \boldsymbol{p} \cdot \overline{\boldsymbol{r}}}
			\hat{\psi}_\sigma^\dagger \left(  \boldsymbol{r} + \frac{\overline{\boldsymbol{r}}}{2} , t \right)
			\hat{\psi}_\sigma \left(  \boldsymbol{r} - \frac{\overline{\boldsymbol{r}} }{2} , t \right)
	\label{def_Wigner_opr_collinear}
\end{gather}
and the two-particle one
\begin{gather}
	\hat{f}_{\sigma \sigma'} (\boldsymbol{r}_1, \boldsymbol{p}_1, \boldsymbol{r}_2, \boldsymbol{p}_2, t)
	\equiv
		\int
		\frac{d^3 \overline{r}_1}{h^3}
		\frac{d^3 \overline{r}_2}{h^3}
			e^{i \boldsymbol{p}_1 \cdot \overline{\boldsymbol{r}}_1 + i \boldsymbol{p}_2 \cdot \overline{\boldsymbol{r}}_2}
			\hat{\psi}_\sigma^\dagger \left(  \boldsymbol{r}_1 + \frac{\overline{\boldsymbol{r}}_1 }{2} , t \right)
			\hat{\psi}_{\sigma'}^\dagger \left(  \boldsymbol{r}_2 + \frac{\overline{\boldsymbol{r}}_2 }{2} , t \right)
			\hat{\psi}_{\sigma'} \left(  \boldsymbol{r}_2 - \frac{\overline{\boldsymbol{r}}_2 }{2} , t \right)
			\hat{\psi}_\sigma \left(  \boldsymbol{r}_1 - \frac{\overline{\boldsymbol{r}}_1 }{2} , t \right)
	,
	\label{def_Wigner_opr_collinear_2}
\end{gather}
\end{widetext}
as adopted by Hasegawa and Shimizu.\cite{bib:3889}
$h = 2 \pi$ is the Planck constant.
The Wigner distribution is often interpreted as a quasiprobability distribution\cite{bib:132, bib:3958} in phase space.

The time development of the one-particle Wigner distribution operator is governed by the Heisenberg equation, found in textbooks:
\begin{gather}
	i
	\frac{\partial}{\partial t} \hat{f}_{\sigma} (\boldsymbol{r}, \boldsymbol{p}, t)
	=
		[ \hat{f}_{\sigma} (\boldsymbol{r}, \boldsymbol{p}, t), \hat{H} ]
	.
	\label{EOM_Winger_opr_H}
\end{gather}
By employing the anticommutation relations for $\hat{\psi}_\sigma^\dagger$ and $\hat{\psi}_\sigma$,
the right-hand side of eq. (\ref{EOM_Winger_opr_H}) can be basically calculated.
One may, however, find soon that the vector  potential in the kinetic-energy operator is disturbing for
obtaining the expression of the Wigner distribution operator.
Therefore we introduce an approximation in which the vector potential varies so slowly in real space that
its second- and higher-order derivatives vanish everywhere:
\begin{gather}
	\boldsymbol{A}_{\sigma}
	( \boldsymbol{r} + \boldsymbol{r}', t )
	\approx
		\boldsymbol{A}_{\sigma} (\boldsymbol{r}, t)
		+
		( \boldsymbol{r}' \cdot \nabla_{\boldsymbol{r}} )
		\boldsymbol{A}_{\sigma} (\boldsymbol{r}, t)
	\label{slow_var_approx_for_vec_pot}
\end{gather}
for arbitrary $\boldsymbol{r}$ and $\boldsymbol{r}'$.

Since the vector potential exists in the present system,
it is appropriate to perform the variable transformation
\begin{gather}
	(\boldsymbol{r}, \boldsymbol{p}, t)
	\to
	(\widetilde{\boldsymbol{r}} \equiv \boldsymbol{r}, \boldsymbol{\Pi} \equiv \boldsymbol{p} + \boldsymbol{A}_{\sigma} (\boldsymbol{r}, t)/c, \widetilde{t} \equiv t )
	,
	\label{variable_transform}
\end{gather}
as pointed out by Kubo.\cite{bib:4079}
The physical quantities should be calculated not via the canonical momentum $\boldsymbol{p}$,
but via the physical (gauge covariant) momentum $\boldsymbol{\Pi}$.
We omit the spin index for $\boldsymbol{\Pi}$ since the vector potential actually does not have spin dependence.
The function form of the one-particle Wigner distribution operator can change via the transformation as
\begin{gather}
	\hat{f}_\sigma (\boldsymbol{r}, \boldsymbol{p}, t)
	\to
	\hat{f}_\sigma (\widetilde{\boldsymbol{r}}, \boldsymbol{\Pi} - \boldsymbol{A}_{\sigma} (\boldsymbol{r}, t)/c , \widetilde{t})
	\equiv
		\hat{\widetilde{f}}_\sigma (\widetilde{\boldsymbol{r}}, \boldsymbol{\Pi} , \widetilde{t} )
	,
	\label{replace_p_by_pi}
\end{gather}
for which we can derive the equation of motion (EOM) for the one-particle WDF
$
\widetilde{f}_{\sigma} (\widetilde{\boldsymbol{r}}, \boldsymbol{\Pi}, \widetilde{t})
\equiv
\langle \hat{\widetilde{f}}_{\sigma} (\widetilde{\boldsymbol{r}}, \boldsymbol{\Pi}, \widetilde{t}) \rangle
$
as (see Appendix for the details of derivation)
\begin{widetext}
\begin{gather}
		i
		\frac{\partial \widetilde{f}_\sigma}{\partial \widetilde{t}} 
		+
		\frac{i}{c}
		\frac{\partial \boldsymbol{A}_{\sigma}}{\partial \widetilde{t}}
		\cdot
		\nabla_{\boldsymbol{\Pi}} 
		\widetilde{f}_{\sigma}
	=
		-\frac{i}{m}
		\boldsymbol{\Pi}
		\cdot
		\nabla_{\widetilde{\boldsymbol{r}}}
		\widetilde{f}_\sigma
		+
		\frac{i}{m c}	
		(\boldsymbol{\Pi} \times \boldsymbol{B}_{\sigma})
		\cdot
		\nabla_{\boldsymbol{\Pi}}
		\widetilde{f}_\sigma
		-
		\int
			\frac{d^3 \overline{r}}{h^3}
			d^3 \Pi' \,
			e^{i ( \boldsymbol{\Pi} - \boldsymbol{\Pi}' ) \cdot \overline{\boldsymbol{r}}}
			V_{\sigma}^{\mathrm{(W)}} \left(  \widetilde{\boldsymbol{r}} , \frac{\overline{\boldsymbol{r}}}{2} , \widetilde{t} \right)
			\widetilde{f}_{\sigma} (\widetilde{\boldsymbol{r}}, \boldsymbol{\Pi}', \widetilde{t})
	\nonumber \\		
		-
		\int
		\frac{d^3 \overline{r}}{h^3}
		d^3 r'
		d^3 \Pi_1
		d^3 \Pi_2 \,
			e^{i ( \boldsymbol{\Pi} - \boldsymbol{\Pi}_1  ) \cdot \overline{\boldsymbol{r}}}
			v^{\mathrm{(W)}} \left( \widetilde{\boldsymbol{r}}  - \boldsymbol{r}', \frac{\overline{\boldsymbol{r}}}{2} \right)
			\sum_{\sigma'}
			\widetilde{f}_{\sigma \sigma'} (\widetilde{\boldsymbol{r}}, \boldsymbol{\Pi}_1, \boldsymbol{r}', \boldsymbol{\Pi}_2, \widetilde{t})
	.
	\label{EOM_WDF1_with_vec_pot}
\end{gather}
\end{widetext}
We omit the tildes on the symbols in what follows
since the removal of them would not cause confusion for the reader.

\subsection{Linear-response functions in STLS approach}

The equal-time pair correlation function in the ground state is defined as
\begin{gather}
	g_{\sigma \sigma'} (\boldsymbol{r}, \boldsymbol{r}')
	\equiv
		\frac{n^{(0)}_{\sigma \sigma'} (\boldsymbol{r}, \boldsymbol{r}')}{ n^{(0)}_{\sigma} (\boldsymbol{r}) n^{(0)}_{\sigma'} (\boldsymbol{r}')  }
	,
	\label{def_pair_corr_func}
\end{gather}
where
$n_{\sigma}^{(0)}$ is the electron density of spin $\sigma$
and
$n^{(0)}_{\sigma \sigma'} (\boldsymbol{r}, \boldsymbol{r}')$
is the equal-time two-electron distribution in the ground state.
To obtain the EOM for the one-particle WDF in a closed form,
we introduce the following ansatz for an inhomogeneous system:\cite{bib:4075, bib:4076}
\begin{gather}
	f_{\sigma \sigma'} (\boldsymbol{r}, \boldsymbol{\Pi}, \boldsymbol{r}', \boldsymbol{\Pi}', t)
	\approx
		f_{\sigma} (\boldsymbol{r}, \boldsymbol{\Pi},  t)
		g_{\sigma \sigma'} (\boldsymbol{r},  \boldsymbol{r}')
		f_{\sigma'} (\boldsymbol{r}', \boldsymbol{\Pi}',  t)
	.
	\label{approx_WDF2_using_pair_corr_vec_pot}
\end{gather}
This factorization, which we call the STLS approximation in what follows,
assumes that the correlation effects in the perturbed system is represented accurately with $g_{\sigma \sigma'}$, independent of momenta.
This ansatz is the same as for the homogeneous cases
in the original semi-classical\cite{bib:3818}
and the quantum\cite{bib:3889} STLS approaches,
differing only in that $g_{\sigma \sigma'}$ can depend on two position variables separately due to the inhomogeneity.
The decomposition
$
f_{\sigma} (\boldsymbol{r}, \boldsymbol{\Pi},  t)
=
f_{\sigma}^{(0)} (\boldsymbol{r}, \boldsymbol{\Pi})
+
f_{\sigma}^{(1)} (\boldsymbol{r}, \boldsymbol{\Pi},  t)
$
of the one-particle WDF after the perturbation is turned on
into its deviation $f_{\sigma}^{(1)}$ from the unperturbed one $f_{\sigma}^{(0)}$
for eq. (\ref{EOM_WDF1_with_vec_pot})
together with eq. (\ref{approx_WDF2_using_pair_corr_vec_pot})
allows us to get the EOM for $f_{\sigma}^{(1)}$ in the linear-response regime.

It is known that one can obtain the distribution for position variable
by integrating out the momentum variable in the one-particle WDF.\cite{bib:132, bib:3958}
From the scalar response of the one-particle WDF in frequency domain to the perturbation potential 
\begin{gather}
	F_{\sigma \sigma'} (\boldsymbol{r}, \boldsymbol{\Pi}, \boldsymbol{r}', \omega)
	\equiv
		\frac{\delta f_{\sigma}^{(1)} (\boldsymbol{r}, \boldsymbol{\Pi}, \omega) }{\delta V_{\mathrm{ext} \sigma'} (\boldsymbol{r}', \omega)}
	\label{vec_repr_resp_WDF_r_pi_omega_to_ext}
\end{gather}
and the vector response to the perturbation vector potential
\begin{gather}
	\boldsymbol{F}_{\sigma \sigma'} (\boldsymbol{r}, \boldsymbol{\Pi}, \boldsymbol{r}', \omega)
	\equiv
		\frac{\delta f_{\sigma}^{(1)} (\boldsymbol{r}, \boldsymbol{\Pi}, \omega) }{\delta \boldsymbol{A}_{\mathrm{ext} \sigma'} (\boldsymbol{r}', \omega)}
	,
	\label{vec_repr_resp_WDF_r_pi_omega_to_ext_vec_pot}
\end{gather}
we can thus calculate the (electron) density response function
as the functional derivative of induced electron density with respect to $V_{\mathrm{ext}}$
\begin{gather}
	\chi_{\sigma \sigma'} (\boldsymbol{r}, \boldsymbol{r}', \omega )
	=
		\frac{ \delta n_{\mathrm{ind} \sigma} (\boldsymbol{r}, \omega) }{\delta V_{\mathrm{ext} \sigma'} (\boldsymbol{r}', \omega) }
	\nonumber \\
	=
		\int d^3 \Pi \,
			F_{\sigma \sigma'} (\boldsymbol{r}, \boldsymbol{\Pi}, \boldsymbol{r}', \omega)
	\label{resp_of_el_dens_to_W}
\end{gather}
and the current density response function
as the functional derivative of induced current density with respect to $\boldsymbol{A}_{\mathrm{ext}}$
\begin{gather}
	\chi_{\sigma j, \sigma' j'} (\boldsymbol{r}, \boldsymbol{r}', \omega )
	=
		\frac{ \delta j_{\mathrm{ind} \sigma j} (\boldsymbol{r}, \omega) }{\delta A_{\mathrm{ext} \sigma' j'} (\boldsymbol{r}', \omega) }
	\nonumber \\
	=
		-
		\frac{1}{m}
		\int d^3 \Pi \,
			\Pi_j
			F_{\sigma \sigma' j'} (\boldsymbol{r}, \boldsymbol{\Pi}, \boldsymbol{r}', \omega)
	,
\end{gather}
where the negative sign comes from the negative charge of an electron.
These two response functions are of particular importance compared to those  related to other quantities
since the electron density and the current density are the fundamental quantities in (TD)DFT.\cite{bib:3198, bib:4084}
The response functions of other physical quantities, such as an electric dipole or an orbital angular momentum,
could also be defined and calculated in similar ways,
provided that the order of operators and corresponding Wigner representation\cite{bib:4079} in classical variables are taken care of.

What is needed for us is now the EOM for the response of the one-particle WDF within the STLS approximation.
By using the functional derivatives
\begin{gather}
	\frac{
		\delta
		V_{\mathrm{ext} \sigma} \left(  \boldsymbol{r} \pm \frac{\overline{\boldsymbol{r}}}{2} , \omega \right)
	}{\delta V_{\mathrm{ext} \sigma'} (\boldsymbol{r}', \omega)}
	=
		2
		\delta_{\sigma \sigma'}
		\delta (\overline{\boldsymbol{r}} \mp 2 (\boldsymbol{r}' - \boldsymbol{r}) )
\end{gather}
and
\begin{widetext}
\begin{gather}
	\frac{\delta}{\delta \boldsymbol{A}_{\mathrm{ext} \sigma'} (\boldsymbol{r}', \omega)}
	\Bigg[
		-\frac{\omega}{c} \boldsymbol{A}_{\mathrm{ext} \sigma} (\boldsymbol{r}, \omega)
		+
		i
		\frac{\boldsymbol{\Pi}}{m} \times \frac{\boldsymbol{B}_{\mathrm{ext} \sigma} (\boldsymbol{r}, \omega)}{c}
	\Bigg]
	\cdot
	\nabla_{\boldsymbol{\Pi}}
	f_\sigma^{(0)} (\boldsymbol{r}, \boldsymbol{\Pi} )
	=
	\nonumber \\
		-
		\delta_{\sigma \sigma'}
		\frac{\omega}{c}
		\delta (\boldsymbol{r} - \boldsymbol{r}')
		\nabla_{\boldsymbol{\Pi}}
		f_\sigma^{(0)} (\boldsymbol{r}, \boldsymbol{\Pi} )
		+
		\frac{i}{m c}
		\delta_{\sigma \sigma'}
		[ \nabla_{\boldsymbol{r}}	\delta (\boldsymbol{r} - \boldsymbol{r}')	]
		\times
		[ \boldsymbol{\Pi} \times \nabla_{\boldsymbol{\Pi}} f_\sigma^{(0)} (\boldsymbol{r}, \boldsymbol{\Pi} ) ]
	,
\end{gather}
we are able to get the equation we want by taking the functional derivatives of the EOM for $f_{\sigma}^{(1)}$ with respect to the perturbation as
\begin{gather}
	\Bigg[
		\omega
		+
		i
		\frac{\boldsymbol{\Pi}}{m}
		\cdot
		\nabla_{\boldsymbol{r}}
		-
		i
		\frac{\boldsymbol{\Pi}}{m} \times \frac{\boldsymbol{B}_{0 \sigma} (\boldsymbol{r})}{c}
		\cdot
		\nabla_{\boldsymbol{\Pi}}
		+
		V_{0 \sigma}^{\mathrm{(W)}} \left( \boldsymbol{r} , \frac{\nabla_{\boldsymbol{\Pi}}}{2 i} \right)
	\Bigg]
	\begin{pmatrix}
		F_{\sigma \sigma'}^{\mathrm{STLS}} (\boldsymbol{r}, \boldsymbol{\Pi}, \boldsymbol{r}', \omega) \\
		\boldsymbol{F}_{\sigma \sigma'}^{\mathrm{STLS}} (\boldsymbol{r}, \boldsymbol{\Pi}, \boldsymbol{r}', \omega) \\
	\end{pmatrix}
	-
	\begin{pmatrix}
		\mathcal{C}_{\sigma \sigma'} [ F^{\mathrm{STLS}} ] (\boldsymbol{r}, \boldsymbol{\Pi}, \boldsymbol{r}', \omega) \\
		\mathcal{C}_{\sigma \sigma'} [ \boldsymbol{F}^{\mathrm{STLS}} ] (\boldsymbol{r}, \boldsymbol{\Pi}, \boldsymbol{r}', \omega) \\
	\end{pmatrix}
	=
	\nonumber \\
	\delta_{\sigma \sigma'}
	\begin{pmatrix}
		s_{\sigma} (\boldsymbol{r}, \boldsymbol{\Pi}, \boldsymbol{r}')
		\\
		\boldsymbol{s}_{\sigma} (\boldsymbol{r}, \boldsymbol{\Pi}, \boldsymbol{r}')
	\end{pmatrix}
	,
	\label{resp_WDF_r_pi_omega_to_ext_pot_and_vec_pot}
\end{gather}
where
the source term for the scalar response
\begin{gather}
		s_{\sigma} (\boldsymbol{r}, \boldsymbol{\Pi}, \boldsymbol{r}')
		\equiv
			2
			\int
			\frac{d^3 \Pi'}{h^3} 
			[
				- e^{2 i ( \boldsymbol{\Pi} - \boldsymbol{\Pi}' ) \cdot (\boldsymbol{r}' - \boldsymbol{r}) }
				+ e^{-2 i ( \boldsymbol{\Pi} - \boldsymbol{\Pi}' ) \cdot (\boldsymbol{r}' - \boldsymbol{r}) }
			]
			f_{\sigma}^{(0)} (\boldsymbol{r}, \boldsymbol{\Pi}')
	\label{def_scalar_source_in_EOM}
\end{gather}
and
that for the vector one
\begin{gather}
		\boldsymbol{s}_{\sigma} (\boldsymbol{r}, \boldsymbol{\Pi}, \boldsymbol{r}')
		\equiv
			-
			\frac{\omega}{c}
			\delta (\boldsymbol{r} - \boldsymbol{r}')
			\nabla_{\boldsymbol{\Pi}}
			f_\sigma^{(0)} (\boldsymbol{r}, \boldsymbol{\Pi} )
			+
			\frac{i}{m c}
			[ \nabla_{\boldsymbol{r}}	\delta (\boldsymbol{r} - \boldsymbol{r}')	]
			\times
			[ \boldsymbol{\Pi} \times \nabla_{\boldsymbol{\Pi}}
			f_\sigma^{(0)} (\boldsymbol{r}, \boldsymbol{\Pi} ) ]
\end{gather}
have been defined.
\begin{gather}
	\mathcal{C}_{\sigma \sigma'} [ h ] (\boldsymbol{r}, \boldsymbol{\Pi}, \boldsymbol{r}', \omega) 
	\equiv
	\nonumber \\
		\int
		d^3 r''
		d^3 \Pi' \,
			v^{\mathrm{(W)}} \left( \boldsymbol{r}  - \boldsymbol{r}'' , \frac{\nabla_{\boldsymbol{\Pi}} }{2 i} \right)
			\sum_{\sigma''}
				[
					f_{\sigma}^{(0)} (\boldsymbol{r}, \boldsymbol{\Pi})
					h_{\sigma'' \sigma'} (\boldsymbol{r}'', \boldsymbol{\Pi}', \boldsymbol{r}', \omega)
					+
					h_{\sigma \sigma'} (\boldsymbol{r}, \boldsymbol{\Pi}, \boldsymbol{r}', \omega)					
					f_{\sigma''}^{(0)} (\boldsymbol{r}'', \boldsymbol{\Pi}')
				]
				g_{\sigma \sigma''} (\boldsymbol{r}, \boldsymbol{r}'')
	\label{def_collision_func}
\end{gather}
\end{widetext}
is the collision functional, which describes the exchange correlation effects within the STLS approximation in addition to the bare interaction.
The linear EOM for the response in eq. (\ref{resp_WDF_r_pi_omega_to_ext_pot_and_vec_pot}) is the main result of the present work,
applicable to perturbation composed of an arbitrary electric field and a slowly-varying magnetic field.
It was derived as the central part of the quantum inhomogeneous STLS approach for the first time to the authors' best knowledge,
and is also a natural extension of the approach developed by Hasegawa and Shimizu\cite{bib:3888, bib:3889} for homogeneous systems.
Since the EOM obtained here is non-local,
it has to be solved for the two independent variables $\boldsymbol{r}$ and $\boldsymbol{r}'$ together with $\boldsymbol{\Pi}$ for a fixed $\omega$,
generally leading to higher computational cost than homogeneous cases. 
In the reciprocal-space language, that is equivalently said that the modes in different wave vectors of perturbation are coupled.
When the EOM is solved practically, $\omega$ as the operator in EOM should be replaced by $\omega + i \delta$ using a small positive constant $\delta$ for ensuring causality.

The EOM obtained here contains the two generally unknown ingredients,
the pair correlation function $g_{\sigma \sigma'}$ and the one-particle WDF $f^{(0)}_\sigma$ both in the equilibrium.
The former can be calculated from the interacting two-electron distribution
by using the well known fluctuation dissipation theorem (FDT)\cite{bib:4109}
\begin{gather}
	n_{\sigma}^{(0)} (\boldsymbol{r})
	n_{\sigma'}^{(0)} (\boldsymbol{r}')
	-
	\frac{1}{\pi}
	\int_0^\infty
	d \omega \,
		\mathrm{Im} \,
		\chi_{\sigma \sigma'} (\boldsymbol{r}, \boldsymbol{r}', \omega)
	\nonumber \\
	=
		\delta_{\sigma \sigma'}
		\delta (\boldsymbol{r} - \boldsymbol{r}' )
		n_\sigma^{(0)} (\boldsymbol{r}) 
		+
		n_{\sigma \sigma'}^{(0)} (\boldsymbol{r}, \boldsymbol{r}')
	\label{S_r_t0_from_dens_resp}
\end{gather}
as long as the density response function and the interacting electron density are known.
If one is able to prepare $f^{(0)}_\sigma$ and an initial guess for $g_{\sigma \sigma'}$,
and to solve the EOM, $\chi_{\sigma \sigma'}$ is obtained, from which the new $g_{\sigma \sigma'}$ is calculated via the FDT.
It is then put into the EOM to be solved again to update $\chi_{\sigma \sigma'}$.
This cycle is repeated until it converges similarly to homogeneous cases.
It is noted here that the FDT, which itself is exact, does not give the exact two-electron distribution when used with the EOM due to the STLS-approximated response function.
Such an error can be carried over for the subsequent iterations
even when one has started from the first iteration with the exact two-electron distribution.
After the cycles,
the interaction energy is then calculated as
$
E_{\mathrm{int}} = \sum_{\sigma, \sigma'} \int d^3 r d^3 r' \, n_{\sigma \sigma'}^{(0)} (\boldsymbol{r}, \boldsymbol{r}') v (\boldsymbol{r} - \boldsymbol{r}')/2.
$
Using the FDT and the Hartree energy
$
E_{\mathrm{H}} = \int d^3 r d^3 r' \, n^{(0)} (\boldsymbol{r}) n^{(0)} (\boldsymbol{r}') v (\boldsymbol{r} - \boldsymbol{r}')/2,
$
the difference $E_{\mathrm{int-H}}$ between the interaction and the Hartree energies are written as
\begin{gather}
	E_{\mathrm{int-H}}
	=
		-
		\frac{1}{2}
		\int
		d^3 r
		d^3 r' \,
			v (\boldsymbol{r} - \boldsymbol{r}' )
			\delta (\boldsymbol{r} - \boldsymbol{r}')
			n^{(0)} (\boldsymbol{r})
	\nonumber \\
		-
			\frac{1}{2}
			\int
			d^3 r
			d^3 r' \,
				v (\boldsymbol{r} - \boldsymbol{r}' )
			\sum_{\sigma, \sigma'}
			\int_0^\infty
			\frac{d \omega}{\pi} 
				\mathrm{Im} \,
				\chi_{\sigma \sigma'} (\boldsymbol{r}, \boldsymbol{r}', \omega)
	.
	\label{E_corr_FDT}
\end{gather}
The first term on the right-hand side of the expression above represents the removal of the self-interaction\cite{bib:4109},
which depends only on the total number of electrons.
It can thus be ignored for an electron-number-conserved system.
The expression of correlation energy for a scaled interaction as a version of eq. (\ref{E_corr_FDT}),
called the adiabatic connection and fluctuation dissipation theorem (ACFDT),
is employed for connecting the non-interacting KS and the true (interacting) systems by Dobson et. al.\cite{bib:4075, bib:4076} for their construction of inhomogeneous STLS approach.

Although we need $f_\sigma^{(0)}$ for constructing the EOM,
it is difficult to access even the ground state wave function of an interacting system in general.
In such a case for a practical calculation, we have to use an alternative one-particle WDF instead of the true one.
It is clear now that a self-consistent inhomogeneous STLS calculation can be followed, apart from numerical inaccuracy, by the following three sources of errors:
The first is the STLS approximation.
The second is the usage of an incorrect one-particle WDF (and hence the electron density) for the EOM.
The third is the discrepancy between the converged two-electron distribution and the exact one which intrudes during the self-consistency cycles.
Since the first one cannot be avoided among the three for an STLS calculation,
its validity should be examined first of all.
The many-body wave functions of a generic interacting system are quite difficult to obtain even numerically, however,
we have to resort to detailed analyses on specific systems to draw possible generic lessons.
We will therefore inquire later the correlation effects in an analytically solvable two-electron system,
from which only the effects coming from the STLS approximation can be extracted.

It is well known that the exact KS potential for ground state of an interacting two-electron system can be constructed
if the electron density of the spin-singlet ground state is known.
In such a case, the doubly occupied real KS orbital is obtained from the relation $n^{(0)} (\boldsymbol{r}) = 2 \psi^{\mathrm{KS}} (\boldsymbol{r})^2$,
where $2$ is the spin degeneracy and $\psi^{\mathrm{KS}}$ is the spatial part of the KS orbital.
This relation and the KS equation provide the exact KS potential as
\begin{gather}
	V_{\mathrm{KS}} (\boldsymbol{r})
	=
		\frac{\nabla^2 \sqrt{n^{(0)} (\boldsymbol{r}) } }{2 m \sqrt{n^{(0)} (\boldsymbol{r}) }}
		+
		\mathrm{const.}
	\label{exact_V_KS_for_two_electrons}
\end{gather}
The exact KS potential calculated in this way allows one to analyze the exchange correlation effects on an interacting system without performing a DFT calculation.\cite{bib:3919, bib:3915, bib:3914}
We will use the relation in eq. (\ref{exact_V_KS_for_two_electrons}) later for constructing the KS non-interacting density response function $\chi^{\mathrm{KS}}$ of an interacting two-electron system.
Since there exists a generic relation $f_{\mathrm{xc}} = \chi^{\mathrm{KS} -1} - \chi^{-1} - v$, coming from the definition of KS potential,
one can obtain in principle the exact exchange correlation kernel $f_{\mathrm{xc}}$ for an interacting two-electron system by using its exact $V_{\mathrm{KS}}$.
Thiele and K\"ummel\cite{bib:3998} constructed the exact kernel for a two-electron system by inverting numerically the exact and the KS density response functions
with careful treatment of the inherent singularity contained in the inverse response function.

\subsection{Slow-variation approximation for potential and interaction}

It is interesting to see the form of EOM in eq. (\ref{resp_WDF_r_pi_omega_to_ext_pot_and_vec_pot})
when the slow-variation approximation,
which neglects the second- and higher-order derivatives,
is introduced to the inherent potential and the interaction between electrons. 
In such a case, by remembering the relation $\boldsymbol{E}_{\sigma} (\boldsymbol{r}, t) = \nabla_{\boldsymbol{r}} V (\boldsymbol{r}, t) - c^{-1} \partial \boldsymbol{A}_{\sigma} (\boldsymbol{r}, t)/\partial t$,
the inherent-potential term is approximated as
$
V_{0 \sigma}^{\mathrm{(W)}} \left( \boldsymbol{r} , \nabla_{\boldsymbol{\Pi}} / (2 i) \right)
\approx
-i
\boldsymbol{E}_{0 \sigma} (\boldsymbol{r})
\cdot
\nabla_{\boldsymbol{\Pi}}
$
.
The collision functional is thus rewritten as
\begin{widetext}
\begin{gather}
	\mathcal{C}_{\sigma \sigma'} [ h ] (\boldsymbol{r}, \boldsymbol{\Pi}, \boldsymbol{r}', \omega) 
	\approx
		i
		m^2
		\boldsymbol{E}_{\sigma \sigma'}^{\mathrm{Hxc} } [h] (\boldsymbol{r}, \boldsymbol{r}', \omega)
		\cdot
		\nabla_{\boldsymbol{\Pi}}
		f_{\sigma}^{(0)} (\boldsymbol{r}, \boldsymbol{\Pi})
		+ i
		\boldsymbol{E}_{\sigma}^{\mathrm{Hxc (0)} }  (\boldsymbol{r})  
		\cdot
		\nabla_{\boldsymbol{\Pi}} h_{\sigma \sigma'} (\boldsymbol{r}, \boldsymbol{\Pi}, \boldsymbol{r}', \omega)					
	,
\end{gather}
where
\begin{gather}
	\boldsymbol{E}_{\sigma \sigma'}^{\mathrm{Hxc} } [h] (\boldsymbol{r}, \boldsymbol{r}', \omega)  
	\equiv
			-
			\frac{1}{m^2}
			\int
			d^3 r'' \,
			\nabla_{\boldsymbol{r}}
			v (\boldsymbol{r} - \boldsymbol{r}'')
			\sum_{\sigma''}
			g_{\sigma \sigma''} (\boldsymbol{r}, \boldsymbol{r}'')
				\int
				d^3 \Pi' \,
					h_{\sigma'' \sigma'} (\boldsymbol{r}'', \boldsymbol{\Pi}', \boldsymbol{r}', \omega)
	,
	\label{def_E_H_xc}
	\\
	\boldsymbol{E}_{\sigma}^{\mathrm{Hxc (0)} }  (\boldsymbol{r})
	\equiv
		-
		\int
		d^3 r' \,
		\nabla_{\boldsymbol{r}} v (\boldsymbol{r} - \boldsymbol{r}')
		\sum_{\sigma''}
			n_{\sigma''}^{(0)} (\boldsymbol{r}' )
			g_{\sigma \sigma''} (\boldsymbol{r}, \boldsymbol{r}')
	.
	\label{def_E_H_xc_0}
\end{gather}
\end{widetext}
We introduced the factor $m^{-2}$ in the definition of $\boldsymbol{E}^{\mathrm{Hxc} }$ only for making it have the dimension of an electric field.
$\boldsymbol{E}^{\mathrm{Hxc (0)} }$
and
$\boldsymbol{E}^{\mathrm{Hxc} }$
can be interpreted as the electric fields generated by the unperturbed and perturbed electron densities, respectively.
They contain not only the bare-interaction (Hartree) effects but also the exchange correlation (xc) effects within the STLS approximation.\cite{bib:4076}
Equation (\ref{resp_WDF_r_pi_omega_to_ext_pot_and_vec_pot}) for the present case thus becomes
\begin{widetext}
\begin{gather}
	\Bigg[
		\omega
		+
		i
		\frac{\boldsymbol{\Pi}}{m}
		\cdot
		\nabla_{\boldsymbol{r}}
		+
		i
		\left\{
			-\boldsymbol{E}_{0 \sigma}^{\mathrm{eff}} (\boldsymbol{r})
			-\frac{\boldsymbol{\Pi}}{m} \times \frac{\boldsymbol{B}_{0 \sigma} (\boldsymbol{r})}{c}
		\right\}
		\cdot
		\nabla_{\boldsymbol{\Pi}}
	\Bigg]
	\begin{pmatrix}
		F_{\sigma \sigma'}^{\mathrm{STLS}} (\boldsymbol{r}, \boldsymbol{\Pi}, \boldsymbol{r}') \\
		\boldsymbol{F}_{\sigma \sigma'}^{\mathrm{STLS}} (\boldsymbol{r}, \boldsymbol{\Pi}, \boldsymbol{r}') \\
	\end{pmatrix}
	=
	\begin{pmatrix}
		s_{\sigma \sigma'}^{\mathrm{eff}} (\boldsymbol{r}, \boldsymbol{\Pi}, \boldsymbol{r}', \omega)
		\\
		\boldsymbol{s}_{\sigma \sigma'}^{\mathrm{eff}} (\boldsymbol{r}, \boldsymbol{\Pi}, \boldsymbol{r}', \omega)
	\end{pmatrix}
	,
	\label{resp_WDF_r_pi_omega_to_ext_pot_and_vec_pot_slowly_varying}
\end{gather}
where
$
\boldsymbol{E}_{0 \sigma}^{\mathrm{eff}} (\boldsymbol{r})
\equiv
	\boldsymbol{E}_{0 \sigma} (\boldsymbol{r})
	+
	\boldsymbol{E}_{\sigma }^{\mathrm{Hxc (0)} }  (\boldsymbol{r}) 
$
is the effective inherent electric field and
\begin{gather}
	\begin{pmatrix}
		s_{\sigma \sigma'}^{\mathrm{eff}} (\boldsymbol{r}, \boldsymbol{\Pi}, \boldsymbol{r}', \omega)
		\\
		\boldsymbol{s}_{\sigma \sigma'}^{\mathrm{eff}} (\boldsymbol{r}, \boldsymbol{\Pi}, \boldsymbol{r}', \omega)
	\end{pmatrix}
	\equiv
	\delta_{\sigma \sigma'}
	\begin{pmatrix}
		s_{\sigma} (\boldsymbol{r}, \boldsymbol{\Pi}, \boldsymbol{r}')
		\\
		\boldsymbol{s}_{\sigma} (\boldsymbol{r}, \boldsymbol{\Pi}, \boldsymbol{r}')
	\end{pmatrix}
	+
	i
	m^2
	\begin{pmatrix}
		\boldsymbol{E}_{\sigma \sigma'}^{\mathrm{Hxc} } [F^{\mathrm{STLS}}] (\boldsymbol{r}, \boldsymbol{r}', \omega) \\
		\boldsymbol{E}_{\sigma \sigma'}^{\mathrm{Hxc} } [\boldsymbol{F}^{\mathrm{STLS}}] (\boldsymbol{r}, \boldsymbol{r}', \omega) \\
	\end{pmatrix}
	\cdot
	\nabla_{\boldsymbol{\Pi}} f_{\sigma}^{(0)} (\boldsymbol{r}, \boldsymbol{\Pi})
\end{gather}
\end{widetext}
is the effective source.
This EOM is in a form similar to that of the classical one from which Dobson et al.\cite{bib:4075} started to construct their inhomogeneous STLS approach.
If we see the EOM in semi-classical picture as did by them,
the term enclosed by the curly braces on the left-hand side in eq. (\ref{resp_WDF_r_pi_omega_to_ext_pot_and_vec_pot_slowly_varying}) comes from the effective force acting on an electron in the unperturbed system,
while the source comes from the effective perturbing force [see eq. (10) in Ref. \cite{bib:4075}].
From this standpoint,
we could interpret the interacting system as a system in which
the non-interacting electrons in the inherent electric $\boldsymbol{E}_0^{\mathrm{eff}}$ and magnetic $\boldsymbol{B}_0^{\mathrm{eff}}$ fields are perturbed by the source $s^{\mathrm{eff}}$ and $\boldsymbol{s}^{\mathrm{eff}}$.
It is noted here, however,
that the one-particle WDF involved in the EOM is for the interacting system
and the effective source depends on the response $F^{\mathrm{STLS}}$ and $\boldsymbol{F}^{\mathrm{STLS}}$.
The resemblance between the semi-classical and the quantum EOMs with the slow-variation approximation to the interaction and the inherent potential
is also discussed by Hasegawa and Shimizu\cite{bib:3889} for homogeneous systems.

\section{Application to confined interacting two electrons}

We examine a one-dimensional interacting two-electron system which has been analytically solved by Nagy et. al.\cite{bib:3964, bib:2826, bib:2825},
since it enables us to compare the various quantities within the inhomogeneous STLS approximation
and those calculated exactly from their definitions.
We focus on the comparison between the exact and the STLS-approximated density response functions
since they are directly related to the formulation of inhomogeneous STLS approach.

\subsection{Setup and analytic solutions}

\subsubsection{Hamiltonian and energy eigenstates}

Let the (first-quantized) Hamiltonian of the system
\begin{gather}
	\mathcal{H}
	=
		-\frac{1}{2 m}
		\left(
			\frac{\partial^2}{\partial x_1^2}
			+
			\frac{\partial^2}{\partial x_2^2}
		\right)
	\nonumber \\
		+
		\frac{m \omega_0^2}{2}
		(x_1^2 + x_2^2)
		-
		\frac{m \Lambda \omega_0^2}{2}
		(x_1 - x_2)^2
	,
	\label{harm_harm_Hamiltonian}
\end{gather}
where $\omega_0$ is the positive strength of harmonic confining potential
and the dimensionless parameter $\Lambda$ measures the strength of parabolic repulsion between the two electrons.
$x_j (j = 1, 2)$ is the position of the $j$-th electron in the one-dimensional space. 
We assume that $0 \leqq \Lambda \leqq 1/2$ as in the previous works.\cite{bib:3964, bib:2826, bib:2825}
$V_{0 \sigma} (x) = m \omega_0^2 x^2/2$ 
and
$v (x) = -m \Lambda \omega_0^2 x^2/2$
correspond to those in eqs. (\ref{def_V_STLS}) and (\ref{def_H_int_STLS}) , respectively, for this system.
We do not consider the effects of magnetic fields on this system.
The variable transformation for the center-of-mass coordinate $X \equiv (x_1 + x_2)/2$
and the scaled relative coordinate $x_{\mathrm{r}} \equiv (x_1 - x_2)/\sqrt{2}$
decouples the interacting Hamiltonian into two Hamiltonians for independent harmonic oscillators
$
	\mathcal{H}_\mathrm{c}
	\equiv
		-\frac{1}{2 M}
		\frac{\partial^2}{\partial X^2}
		+
		\frac{M \omega_0^2}{2}
		X^2
$
and
$
	\mathcal{H}_\mathrm{r}
	\equiv
		-\frac{1}{2 m}
		\frac{\partial^2}{\partial x_{\mathrm{r}}^2}
		+
		\frac{m \omega_\mathrm{r}^2}{2}
		x_{\mathrm{r}}^2
		,
$
where $M \equiv 2 m$ and
$
	\omega_\mathrm{r}
	\equiv
		\omega_0
		\sqrt{1 - 2 \Lambda}
	\equiv
		\lambda
		\omega_0
		.
$
The time-independent Schr\"odinger equations for them are easily solved by using the Hermite polynomials $H_n$.
For an $n_{\mathrm{c}} = 0, 1, \dots$,
the energy eigenfunction of $\mathcal{H}_{\mathrm{c}}$ is given by
\begin{gather}
	\psi_{\mathrm{c} n_{\mathrm{c}}} (X)
	=
		(2 M \omega_0)^{1/4}
		C_{n_{\mathrm{c}}}
		e^{-Z^2/4}
		H_{n_{\mathrm{c}}}
		\left(
			\frac{Z}{\sqrt{2}}
		\right)
	,
	\label{harm_harm_sol_cm}
\end{gather}
where $C_n \equiv 1/[ (2 \pi)^{1/4} \sqrt{2^n n !}]$
and $Z \equiv \sqrt{2 M \omega_0} X$ is the dimensionless coordinate.
For an $n_{\mathrm{r}} = 0, 1, \dots$,
the energy eigenfunction of $\mathcal{H}_{\mathrm{r}}$ is given by
\begin{gather}
	\psi_{\mathrm{r} n_{\mathrm{r}}} (x_{\mathrm{r}})
	=
		(2 m \omega_{\mathrm{r}})^{1/4}
		C_{n_{\mathrm{r}}}
		e^{-z_{\mathrm{r}}^2/4}
		H_{n_{\mathrm{r}}}
		\left(
			\frac{z_{\mathrm{r}}}{\sqrt{2}}
		\right)
	,
	\label{harm_harm_sol_rel}
\end{gather}
where $z_{\mathrm{r}} \equiv \sqrt{2 m \omega_{\mathrm{r}}} x_{\mathrm{r}}$ is the dimensionless coordinate.
$\psi_{\mathrm{c} n_{\mathrm{c}}} (X)$ and $\psi_{\mathrm{r} n_{\mathrm{r}}} (x_{\mathrm{r}})$ are normalized so that
their integrals over $X$ and $x_{\mathrm{r}}$, respectively, give unity.
With the quantum numbers $n_{\mathrm{c}}$ and $n_{\mathrm{r}}$,
the energy eigenvalue for the whole system is given simply by the sum of two energy eigenvalues for the two oscillators:
$
	E_{n_\mathrm{c} n_\mathrm{r}} 
	=
		\omega_0  ( n_\mathrm{c} + 1/2 )
		+
		\omega_\mathrm{r} ( n_\mathrm{r} + 1/2 )
	=
		\omega_0  [ n_\mathrm{c} + \lambda n_\mathrm{r} + (1 + \lambda)/2 ]
	.
$
The Fermi statistics forces the two-electron wave functions for the energy eigenstates to be in the following two forms:
\begin{gather}
	\Psi_{n_\mathrm{c} n_\mathrm{r}}^{S = 0, S_z = 0} (x_1, \sigma_1, x_2, \sigma_2)
	=
	\nonumber \\
		2^{1/4}
		\psi_{\mathrm{c} n_\mathrm{c}} (X) 
		\psi_{\mathrm{r} n_\mathrm{r}}^{\mathrm{(even)}} (x_{\mathrm{r}}) 
		\phi^{S = 0, S_z = 0} (\sigma_1, \sigma_2) 
	\label{harm_def_Psi_even}
\end{gather}
and
\begin{gather}
	\Psi_{n_\mathrm{c} n_\mathrm{r}}^{S = 1, S_z} (x_1, \sigma_1, x_2, \sigma_2)
	=
	\nonumber \\
		2^{1/4}
		\psi_{\mathrm{c} n_\mathrm{c}} (X) 
		\psi_{\mathrm{r} n_\mathrm{r} }^{\mathrm{(odd)}} (x_{\mathrm{r}}) 
		\phi^{S = 1, S_z} (\sigma_1, \sigma_2) 
	,
	\label{harm_def_Psi_odd}
\end{gather}
where
$\phi^{S S_z}$ is the normalized spin wave function for the spin angular momentum $S$ and its $z$ component $S_z$ of the whole system.
$\psi_{\mathrm{r} n_\mathrm{r}}^{\mathrm{(even)}}$
and
$\psi_{\mathrm{r} n_\mathrm{r}}^{\mathrm{(odd)}}$
are the energy eigenfunctions $\psi_{\mathrm{r} n_\mathrm{r}}$ with even and odd $n_{\mathrm{r}}$'s, respectively,
while there is no restriction on  $\psi_{\mathrm{c} n_\mathrm{c}}$.
The eigenfunctions $\Psi_{n_\mathrm{c} n_\mathrm{r}}$ given above have been normalized so that their integrals over $x_1, \sigma_1, x_2, \sigma_2$ are equal to $2$,
the number of electrons.
The ground state is the spin-singlet state 
with the lowest energy eigenvalue $E_{00} = \omega_0 (1 + \lambda)/2$.
It is given by
$
\Psi_0 (x_1, \sigma_1, x_2, \sigma_2)
\equiv
	\Psi_{0 0}^{S = 0, S_z = 0} (x_1, \sigma_1, x_2, \sigma_2)
=
\psi_0 (x_1, x_2)
\phi^{S = 0, S_z = 0} (\sigma_1, \sigma_2) 
$
,
where
$
\psi_0 (x_1, x_2)
\equiv
	2^{1/4}
	\psi_{\mathrm{c} 0} (X) 
	\psi_{\mathrm{r} 0} (x_{\mathrm{r}}) 
$
is the normalized spatial part.

\subsubsection{Electron density and pair correlation function}

Having obtained the wave function of the ground state,
the two-electron distribution and the electron density are calculated straightforwardly as
\begin{gather}
	n_{\sigma_1 \sigma_2}^{(0)} (x_1, x_2)
	=
		| \Psi_0 (x_1, \sigma_1, x_2, \sigma_2) |^2
	\nonumber \\
	=
		2 m \omega_0
		(1 - \delta_{\sigma_1 \sigma_2})
		\frac{\sqrt{\lambda}}{2 \pi}
		\exp
		\left[
			- \frac{(z_1 + z_2)^2}{4} - \frac{\lambda (z_1 - z_2)^2}{4}
		\right]
	\label{harm_harm_two_el_gs}
\end{gather}
and
\begin{gather}
	n_\sigma^{(0)} (x)
	=
		\sum_{\sigma_2}
		\int_{-\infty}^\infty
		d x_2 \,
			n_{\sigma \sigma_2}^{(0)} (x, x_2)
	\nonumber \\
	=
		\sqrt{2 m \omega_0}
		\sqrt{\frac{\lambda}{\pi (1 + \lambda)}}		
		\exp
		\left(
			-
			\frac{\lambda}{1 + \lambda}
			z^2
		\right)
	,
	\label{harm_harm_el_dens_gs}
\end{gather}
respectively, where
$
z \equiv \sqrt{2 m \omega_0} x
$
is the dimensionless coordinate.
The pair correlation function is calculated immediately from its definition in eq. (\ref{def_pair_corr_func}) as
\begin{gather}
	g_{\sigma_1 \sigma_2} (x_1, x_2)
	=
		(1 - \delta_{\sigma_1 \sigma_2})
		\frac{1 + \lambda}{2 \sqrt{\lambda} }
	\cdot
	\nonumber \\
	\cdot
		\exp
		\left[
			-
			\frac{(z_1 + z_2)^2}{4}
			-
			\lambda
			\frac{(z_1 - z_2)^2}{4}
			+
			\frac{\lambda}{1 + \lambda}
			(z_1^2 + z_2^2 )
		\right]
	.
	\label{harm_harm_pair_corr}
\end{gather}
This pair correlation function is not bounded from above in the dissociation limit.
It is easily confirmed by increasing the magnitudes of $z_1$ and $z_2$ with keeping $z_1 + z_2$ constant.
Despite that, this pair correlation function in the collision integral safely gives finite values when used in the EOM, as will be seen later.

\subsection{Exact WDFs for the ground state}

\subsubsection{One-particle WDF}

With an integral for the spatial wave functions in eqs. (\ref{harm_harm_sol_cm}) and (\ref{harm_harm_sol_rel}),
\begin{gather}
	\int_{-\infty}^\infty
	d \overline{x} \,
		e^{i p \overline{x}}
		\psi_{\mathrm{c} 0} \left( \frac{\overline{x} + x}{2} \right)^*
		\psi_{\mathrm{r} 0} \left( \frac{\overline{x} - x}{\sqrt{2}} \right)
	\nonumber \\
	=
		\left( \frac{\lambda}{2} \right)^{1/4}
		\sqrt{\frac{8}{1 + \lambda}}
		\exp
		\left[
			-
			\frac{\lambda z^2 + 2 i (1 - \lambda) \widetilde{p} z + 4 \widetilde{p}^2}{2 ( 1 + \lambda) }
		\right]
	,
\end{gather}
where $\widetilde{p} \equiv p / \sqrt{2 m \omega_0}$ is the dimensionless momentum,
substitution of the ground state wave function $\Psi_0$ into the one-dimensional version of eq. (\ref{WDF1_from_two_el_Psi}) leads to
the exact one-particle WDF:
\begin{gather}
	f_\sigma^{(0)} (x, p)
	=
	\nonumber \\
		\frac{\sqrt{2}}{4 \pi}
		\int_{-\infty}^\infty
		d \overline{X} \,
			e^{i p \overline{X}}
			\psi_{\mathrm{c} 0} \left( \frac{\overline{X} + x}{2} \right)^*
			\psi_{\mathrm{r} 0} \left( \frac{\overline{X} - x}{\sqrt{2}} \right)
	\cdot
	\nonumber \\
	\cdot
		\int_{-\infty}^\infty
		d \overline{x} \,
			e^{-i p \overline{x}}
			\psi_{\mathrm{r} 0} \left( \frac{\overline{x} - x}{\sqrt{2}} \right)^*
			\psi_{\mathrm{c} 0} \left( \frac{\overline{x} + x}{2} \right)
	\nonumber \\
	=
		\frac{2 \sqrt{\lambda}}{\pi ( 1 + \lambda) }
		\exp
		\left(
			-
			\frac{\lambda z^2 + 4 \widetilde{p}^2}{1 + \lambda}
		\right)
	.
	\label{harm_harm_WDF1}
\end{gather}
It is easily confirmed that its integral over momentum coincides with the electron density in eq. (\ref{harm_harm_el_dens_gs}).

\subsubsection{Two-particle WDF}

Substitution of the ground state wave function $\Psi_0$ into the one-dimensional version of eq. (\ref{WDF2_from_two_el_Psi}) leads to
\begin{gather}
	f_{\sigma \sigma'}^{(0)} (x_1, p_1, x_2, p_2)
	=
		( 1 - \delta_{\sigma \sigma'} )
		f_{\mathrm{c} 0} (X, P)
		f_{\mathrm{r} 0} (x_{\mathrm{r}}, p)
	,
	\label{harm_harm_WDF2_factorized}
\end{gather}
where the contribution from the center-of-mass motion
\begin{gather}
	f_{\mathrm{c} 0} (X, P)
	\equiv
		\int_{-\infty}^\infty
		\frac{d \overline{X}}{h}
			e^{i P \overline{X}}
			\psi_{\mathrm{c} 0} \left( X + \frac{\overline{X}}{2} \right)^*
			\psi_{\mathrm{c} 0} \left( X - \frac{\overline{X}}{2} \right)
\end{gather}
for the momentum $P \equiv p_1 + p_2$
and that from the relative motion 
\begin{gather}
	f_{\mathrm{r} 0} (x_{\mathrm{r}}, p_{\mathrm{r}})
	\equiv
		\int_{-\infty}^\infty
		\frac{d \overline{x}}{h}
			e^{i p_{\mathrm{r}} \overline{x}}
			\psi_{\mathrm{r} 0} \left( x_{\mathrm{r}} + \frac{\overline{x}}{2} \right)^*
			\psi_{\mathrm{r} 0} \left( x_{\mathrm{r}} - \frac{\overline{x}}{2} \right)
	,
	\label{harm_harm_WDF2_factorized}
\end{gather}
for the momentum $p_{\mathrm{r}} \equiv (p_1 - p_2)/\sqrt{2}$
are decoupled.
These contributions have the same function form
as the one-particle WDF for a single harmonic oscillator,
which has been calculated in the literature.\cite{bib:132}
The contribution from the center-of-mass motion is given by
$
f_{\mathrm{c} 0} (X, P)
=
\exp [ -2 H_\mathrm{c} (X, P) / \omega_0 ]/ \pi
$
,
whose exponent is interestingly expressed via the Hamiltonian for a classical harmonic oscillator
$
H_\mathrm{c} (X, P)
=
\frac{P^2}{2 M}
+
\frac{M \omega_0^2 X^2}{2}
$
.
By substituting the expressions of $f_{\mathrm{c} 0}$ and similarly calculated $f_{\mathrm{r} 0}$
into eq. (\ref{harm_harm_WDF2_factorized}),
we obtain the exact two-particle WDF:
\begin{widetext}
\begin{gather}
	f_{\sigma \sigma'}^{(0)} (x_1, p_1, x_2, p_2)
	=
		\frac{ 1 - \delta_{\sigma \sigma'} }{\pi^2}
		\exp
		\Bigg[
			-
			(\widetilde{p}_1 + \widetilde{p}_2)^2
			-
			\frac{(z_1 + z_2)^2}{4}
			-
			\frac{ (\widetilde{p}_1 - \widetilde{p}_2)^2 }{\lambda}
			-
			\frac{\lambda (z_1 - z_2)^2}{4}
		\Bigg]
	,
	\label{harm_harm_WDF2}
\end{gather}
\end{widetext}
where the dimensionless variables defined above are used.
It is easily confirmed that its integral over momenta coincides with the two-electron distribution in eq. (\ref{harm_harm_two_el_gs}).

\subsubsection{Validity of STLS approximation}

\begin{figure}[htbp]
\begin{center}
\includegraphics[keepaspectratio,width=5.5cm]{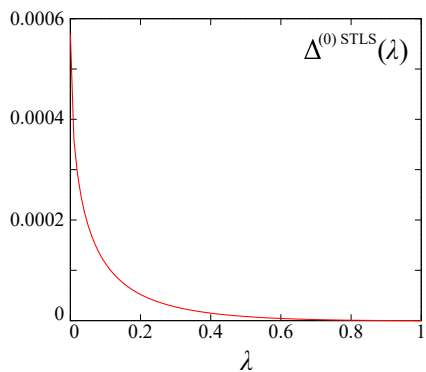}
\end{center}
\caption{
The dimensionless mean squared error per electron in the STLS approximation for the two-particle WDF of the interacting two electrons
as a function of the interaction parameter $\lambda$.
}
\label{Fig_harm_harm_mse}
\end{figure}

Although the STLS approximation is an ansatz for a nonequilibrium two-particle distribution,
it can be helpful to examine its validity for the distributions in equilibrium to get insights into the characteristics of the approximation.
The right-hand side of eq. (\ref{approx_WDF2_using_pair_corr_vec_pot}) in the present case is, from eqs. (\ref{harm_harm_pair_corr}) and (\ref{harm_harm_WDF1}),
\begin{widetext}
\begin{gather}
	f_{\sigma \sigma'}^{(0) \mathrm{STLS}} (x_1, p_1, x_2, p_2)
	\equiv
		f_{\sigma}^{(0)} (x_1, p_1)
		g_{\sigma \sigma'} (x_1, x_2)
		f_{\sigma'}^{(0)} (x_2, p_2)
	\nonumber \\
	=
		(1 - \delta_{\sigma \sigma'})
		\frac{2 \sqrt{\lambda}}{\pi^2 (1 + \lambda)}
		\exp
		\left[
			-\frac{(z_1 + z_2)^2}{4}
			-\frac{\lambda (z_1 - z_2)^2}{4}
			-\frac{4 (\widetilde{p}_1^2 + \widetilde{p}_2^2)}{1 + \lambda}
		\right]
	,
\end{gather}
which clearly differs from the exact two-particle WDF given by eq. (\ref{harm_harm_WDF2})
except for the non-interacting ($\lambda = 1$) case.
The dimensionless mean squared error per electron is calculated analytically as
\begin{gather}
	\Delta^{(0) \mathrm{STLS}} (\lambda)
	\equiv
		\frac{1}{2}
		\int
			d x_1
			d p_1
			d x_2
			d p_2 \,
			| 
				f_{\uparrow \downarrow}^{(0) \mathrm{STLS}} (x_1, p_1, x_2, p_2)
				-
				f_{\uparrow \downarrow}^{(0)} (x_1, p_1, x_2, p_2)
			|^2
			f_{\uparrow \downarrow}^{(0)} (x_1, p_1, x_2, p_2)
	\nonumber \\
	=
		\frac{1}{18 \pi^4}
		+
		\frac{2 \lambda}{3 \pi^4 (1 + \lambda) \sqrt{(5 + \lambda) (1 + 5 \lambda)}}
		-
		\frac{1}{3 \pi^4}
		\sqrt{\frac{\lambda}{(2 + \lambda) (1 + 2 \lambda)}}
	,
\end{gather}
\end{widetext}
plotted in Fig. \ref{Fig_harm_harm_mse}.
We can observe in the figure that the stronger interaction (smaller $\lambda$) leads to the larger error as a monotonic function of $\lambda$.
This result is consistent with generic but naive speculation that the STLS approximation becomes worse as interaction becomes stronger.
Further examinations on whether this tendency is also the case for nonequilibrium and/or infinite systems will be needed in the future.

\subsection{Exact response function}

Since we know all the energy eigenfunction of the present system,
we can calculate the exact density response function from the expression in eq. (\ref{dens_reps_func_two_el}).
The contribution from the transition from the ground state to the excited state having the quantum numbers $n_{\mathrm{c}}$ and $n_{\mathrm{r}}$ 
\begin{gather}
	\chi_{n_\mathrm{c} n_\mathrm{r}} (x, x', \omega)
	\equiv
		\frac{1}{4}
		\Bigg[
			\frac{
				P_{ (0, 0) (n_\mathrm{c}, n_\mathrm{r})} ( x )
				P_{ (0, 0) (n_\mathrm{c}, n_\mathrm{r})} ( x' )^*
			}{\omega - ( E_{n_\mathrm{c} n_\mathrm{r}} - E_{0 0} ) + i \delta}
	\nonumber \\
			-
			\frac{
				P_{ (0, 0) (n_\mathrm{c}, n_\mathrm{r})} ( x )^*
				P_{ (0, 0) (n_\mathrm{c}, n_\mathrm{r})} ( x' )
			}{\omega + ( E_{n_\mathrm{c} n_\mathrm{r}} - E_{0 0} ) + i \delta}
		\Bigg]
	,
	\label{harm_harm_dens_resp_nc_nr}
\end{gather}
where the factor $1/4$ comes from the spin wave function,
is calculated from the transition amplitudes for spatial parts.
Specifically, by using eqs. (\ref{harm_harm_sol_cm}) and (\ref{harm_harm_sol_rel}) for eq. (\ref{def_P_for_dens_resp_two_el}),
we obtain
\begin{gather}
	P_{(0, 0) ( n_\mathrm{c}, n_\mathrm{r} ) } (x)
	=
		\frac{\sqrt{2 m \omega_0} (-1)^{n_\mathrm{r}} 2 \lambda^{(n_\mathrm{c} + 1) /2}}{ (1 + \lambda)^{( n_\mathrm{c} + n_\mathrm{r} + 1 )/2} \sqrt{ 2^{n_\mathrm{c} + n_\mathrm{r}}  n_\mathrm{c} ! n_\mathrm{r} ! \pi }}
	\cdot
	\nonumber \\
	\cdot
		\exp \left( -\frac{\lambda}{1 + \lambda} z^2 \right)
		H_{n_\mathrm{c} + n_\mathrm{r}} \left( \sqrt{\frac{\lambda}{1 + \lambda}} z \right)
	,
	\label{harm_harm_trans_ampl_spatial}
\end{gather}
where we used the integral formula in eq. (\ref{integ_exp_exp_H_H}).
Using the contribution from the transitions to the spin-singlet states
\begin{gather}
	\chi^{(\mathrm{even})} (x, x', \omega)
	\equiv
		\sum_{n_\mathrm{c} = 0}^\infty
		\sum_{ \substack{ \mathrm{even} \, n_\mathrm{r},  \\ (n_\mathrm{c}, n_\mathrm{r}) \ne (0,0)} }^\infty
			\chi_{n_\mathrm{c} n_\mathrm{r}} (x, x', \omega)
	\label{harm_harm_dens_resp_even}
\end{gather}
and that to the spin-triplet states
\begin{gather}
	\chi^{(\mathrm{odd})} (x, x', \omega)
	\equiv
		\sum_{n_\mathrm{c} = 0}^\infty
		\sum_{\mathrm{odd} \, n_\mathrm{r}}^\infty
			\chi_{n_\mathrm{c} n_\mathrm{r}} (x, x', \omega)
	,
\end{gather}
the spin-dependent density response function can be written as
$
\chi_{\uparrow \uparrow}
=
\chi_{\downarrow \downarrow}
=
\chi^{(\mathrm{even})}
+
\chi^{(\mathrm{odd})}
$
and
$
\chi_{\uparrow \downarrow}
=
\chi_{\downarrow \uparrow}
=
\chi^{(\mathrm{even})}
-
\chi^{(\mathrm{odd})}
$
.
Remembering that the perturbation potential is spin independent,
the density response function of total electron density is given by
$\chi = \sum_{\sigma, \sigma'} \chi_{\sigma \sigma'} = 4 \chi^{(\mathrm{even})}$.
The explicit expression for the even part is obtained
by substituting eqs. (\ref{harm_harm_dens_resp_nc_nr}) and (\ref{harm_harm_trans_ampl_spatial}) into eq. (\ref{harm_harm_dens_resp_even}) as
\begin{widetext}
\begin{gather}
	\chi^{\mathrm{(even)}} (x, x', \omega)
	=
		\frac{2 m \omega_0 \lambda}{\pi}
		\exp \left[ -\frac{\lambda}{1 + \lambda} (z^2 + z'^2) \right]
		\sum_{n_\mathrm{c} = 0}^\infty
		\sum_{\mathrm{even} \, n_\mathrm{r}}^\infty
			\frac{\lambda^{n_\mathrm{c}}}{ 2^{n_\mathrm{c} + n_\mathrm{r}}  n_\mathrm{c} ! n_\mathrm{r} !  (1 + \lambda)^{ n_\mathrm{c} + n_\mathrm{r} + 1 } }
	\cdot
	\nonumber \\
	\cdot
			H_{n_\mathrm{c} + n_\mathrm{r}} \left( \sqrt{\frac{\lambda}{1 + \lambda}} z  \right)
			H_{n_\mathrm{c} + n_\mathrm{r}} \left( \sqrt{\frac{\lambda}{1 + \lambda}} z' \right)
			\left[
				\frac{1}{\omega - ( n_\mathrm{c} + \lambda n_\mathrm{r} ) \omega_0 + i \delta}
				-
				\frac{1}{\omega + ( n_\mathrm{c} + \lambda n_\mathrm{r} ) \omega_0 + i \delta}
			\right]
	.
	\label{harm_harm_chi_even_nc_nr}
\end{gather}
\end{widetext}
That for the odd part is also obtained similarly.

\subsection{KS response function}

Putting the electron density in eq. (\ref{harm_harm_el_dens_gs}) into eq. (\ref{exact_V_KS_for_two_electrons}),
we can construct the exact KS potential for the ground state as
\begin{gather}
	V_{\mathrm{KS}} (x)
	=
		\omega_0
		\left( \frac{\lambda}{1 + \lambda} \right)^2
		z^2
		+
		\omega_0
		\frac{(1 - \lambda)^2}{4 (1 + \lambda)}
	.
	\label{harm_harm_V_KS}
\end{gather}
It is clear from this definition that the KS Hamiltonian for the effective non-interacting system is obtained
apart from a constant simply by replacing
$\omega_0$ with $2 \omega_0 \lambda / (1 + \lambda) \equiv \alpha^2 \omega_0$ and 
removing the interaction term in the original Hamiltonian in eq. (\ref{harm_harm_Hamiltonian}).
The constant on the right-hand side in eq. (\ref{harm_harm_V_KS})
was set so that the total energy of the ground state for the KS system coincide with that for the original interacting system.

\begin{figure}
\includegraphics[keepaspectratio,width=5.5cm]{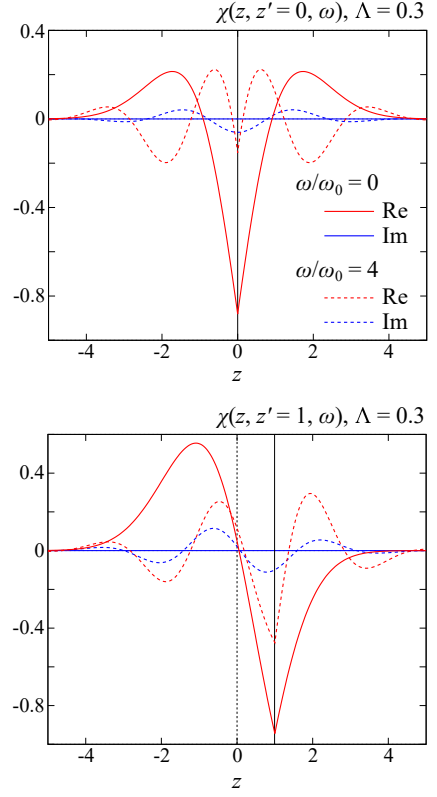}
\caption{
KS total-density response function $\chi^{\mathrm{KS}}$ of the interacting two-electron system for perturbation at dimensionless coordinates $z' = 0$ (top) and $1$ (bottom)
are calculated from eq. (\ref{harm_harm_dens_resp_non_int_para}) and plotted.
The vertical solid lines represent the positions at which the perturbation is applied.
}
\label{Fig_harm_chi_KS}
\end{figure}

The one-particle orbitals for the KS system are those for a harmonic oscillator with its frequency $\alpha^2 \omega_0$.
By substituting these expressions into eq. (\ref{harm_harm_dens_resp_non_int_as_sum})
and using the formula\cite{bib:3963} which expresses the infinite summation for Hermite polynomials as
the products of the parabolic cylinder functions $D_\nu$\cite{bib:Gradshteyn_and_Ryzhik, bib:3956},
we can express the KS total-density response function in a simple form:
\begin{gather}
	\chi^{\mathrm{KS}} (x, x', \omega)
	=
		-
		\frac{2 m}{\pi}
		e^{-\alpha^2 (z^2 + z'^2) /4}
	\cdot
	\nonumber \\
	\cdot
		[
			\Gamma (-\nu)
			D_{\nu} ( \alpha z_>)
			D_{\nu} (-\alpha z_<)
	\nonumber \\
			+
			\Gamma (\nu)
			D_{-\nu} ( \alpha z_>)
			D_{-\nu} (-\alpha z_<)
		]_{ \nu = \omega/(\alpha^2 \omega_0) + i \delta  }
	,
	\label{harm_harm_dens_resp_non_int_para}
\end{gather}
where $z \equiv \sqrt{2 m \omega_0} x$.
$z_<$ and $z_>$ are the lesser and the greater of $z$ and $z'$, respectively.
$\chi^{\mathrm{KS}}$ for some combinations of $\Lambda, z'$, and  $\omega$ were calculated from eq. (\ref{harm_harm_dens_resp_non_int_para}) and plotted in Fig. \ref{Fig_harm_chi_KS}.
We used $\delta  = 0.1$ here and below.
It is clearly seen that the real part of $\chi^{\mathrm{KS}}$ as a function of $z$ can have a cusp at $z = z'$.
Since it is likely that the exact response function $\chi$ has cusps as well,
we should keep in mind for a practical calculation that
the summation over a finite number of polynomials in eq. (\ref{harm_harm_chi_even_nc_nr}) suffers from slow convergence near the cusps.
It is known that the recurrence relation for the derivative of $D_\nu$ similar to that of the Hermite polynomial, eq. (\ref{deriv_Hermite}), holds.
For the $\chi^{\mathrm{KS}}$'s as the functions of $z$ in the figure,
we calculated their $z$ derivatives analytically using this relation
and found that their imaginary parts are continuous at the $z$'s where their real parts are discontinuous in contrast, leading to the real cusps.
These results imply that the existence of cusps is not obvious from the function form of eq. (\ref{harm_harm_dens_resp_non_int_para}).


\subsection{STLS response function}

\subsubsection{Response of one-particle WDF}

The inherent-potential term on the left-hand side of eq. (\ref{resp_WDF_r_pi_omega_to_ext_pot_and_vec_pot}) for this case is
\begin{gather}
	V_{0 \sigma}^{\mathrm{(W)}} \left( x , \frac{1}{2 i} \frac{\partial}{\partial p} \right)
	F_{\sigma \sigma'}^{\mathrm{STLS}} (x, p, x', \omega)
	=
	\nonumber \\
		-i
		\frac{\omega_0 z}{2}
		\frac{\partial F_{\sigma \sigma'}^{\mathrm{STLS}} (x, p, x', \omega)}{\partial \widetilde{p}}
	.
	\label{harm_harm_V_0_in_EOM}
\end{gather}
The source term in eq. (\ref{def_scalar_source_in_EOM}) for this case is
\begin{gather}
	s_{\sigma} (x, p, x')
		=
		\sqrt{\frac{2 m \omega_0 \lambda}{\pi^3 (1 + \lambda) }}
		[
			- e^{2 i \widetilde{p} (z' - z) }
			+ e^{-2 i \widetilde{p} (z' - z) }		
		]
	\cdot
	\nonumber \\
	\cdot
		\exp
		\left[
			-
			\frac{1 + \lambda}{4}
			(z' - z)^2
			-
			\frac{\lambda}{1 + \lambda}
			z^2
		\right]
	,
	\label{harm_harm_source_in_EOM}
\end{gather}
where we used eq. (\ref{harm_harm_WDF1}).
To make the EOM into a one in a numerically tractable form,
we expand the scalar response in the Hermite polynomials as
\begin{gather}
	F_{\sigma \sigma'}^{\mathrm{STLS}} (x, p, x', \omega)
	=
		\sqrt{\frac{2 m}{\omega_0}}
		\frac{e^{-z^2/2 - \widetilde{p}^2}}{\pi}
	\cdot
	\nonumber \\
	\cdot
		\sum_{n, n'}
			c_{\sigma n n'} (\sigma', x', \omega)
			H_n \left( \frac{z}{\sqrt{2}} \right)
			H_{n'} (\widetilde{p})
	.
	\label{harm_harm_resp_in_hermite}		
\end{gather}
Thanks to the relation in eq. (\ref{deriv_Hermite}),
the derivatives of $F^{\mathrm{STLS}}$ with respect to $x$ and $p$ act as the shift in indices of polynomials.
Putting the exact $g_{\sigma \sigma'}$ and $f_\sigma^{(0)}$ calculated above into eq. (\ref{resp_WDF_r_pi_omega_to_ext_pot_and_vec_pot})
and multiplying both sides by $H_n (z/\sqrt{2}) H_{n'} (\widetilde{p})$,
the integration over $x$ and $p$ with the orthogonality of the Hermite polynomials
and tedious manipulations lead to the following linear equation for the expansion coefficients for given $\sigma', x'$, and $\omega$:
\begin{widetext}
\begin{gather}
	i
	\frac{\omega}{\omega_0}
	c_{\sigma n n'}
	+
	\frac{\sqrt{2}}{4}
	c_{\sigma n - 1, n' - 1}
	+
	\sqrt{2}
	(n' + 1)
	c_{\sigma n - 1, n' + 1}
		-
		\frac{\sqrt{2}}{2}
		(n + 1)
		c_{\sigma n + 1, n' - 1}
	\nonumber \\
		+
		\Lambda
		I_{n'}^{(1)}
		\sum_{n_1 = 0}^\infty
			M_{n n_1}
			c_{-\sigma n_1 0}
		+
		\Lambda
		\frac{\sqrt{2}}{1 + \lambda}
		\sum_{n_1 = 0}^\infty
			I_{n n_1}^{(2)}
			c_{\sigma n_1, n' - 1}
	=
		\delta_{\sigma \sigma'}
		B_{n n'} (z')
	,
	\label{harm_harm_resp_coeff_recurr_exact_twoel}
\end{gather}
\end{widetext}
where we regard an expansion coefficient having a polynomial index smaller than $0$ to vanish.
The expressions of the integrals in the equation above are given in Appendix.
Since this equation is an infinite-dimensional matrix equation,
we have to truncate the expansion at a sufficiently large order in a practical calculation
to reduce the "edge effects" of the truncated matrix.
Only for the non-interacting case, the equation can be rewritten to another form that permits one to calculate the expansion coefficients accurately up to an arbitrarily high order (see Appendix).

We confirmed that the non-interacting $\chi^{\mathrm{non-int}}$ calculated from the Lehmann representation in eq. (\ref{harm_harm_dens_resp_non_int_as_sum}) and
that from the EOM in eq. (\ref{harm_harm_resp_coeff_recurr_exact_twoel}),
which gives the exact scalar response for the non-interacting case,
coincide with each other for various combinations of the parameters.
These results corroborate the validity of our quantum inhomogeneous STLS approach.

\begin{figure*}[p]
\caption{
For the interacting two-electron system,
the STLS-approximated $F^{\mathrm{STLS}}$ and the non-interacting $F^{\mathrm{non-int}}$ scalar response of
one-particle WDF for spin-down perturbation applied at dimensionless positions $z' = 0$ and $1$ are plotted.
$\mathrm{Im} F_{\sigma \sigma'}$ for $\omega = 0$ and
$F_{\uparrow \downarrow}^{\mathrm{non-int}}$ are not shown since they vanish.
$m$ and $\omega_0$ are set to unity for the plot.
}
\includegraphics[keepaspectratio,width=16cm]{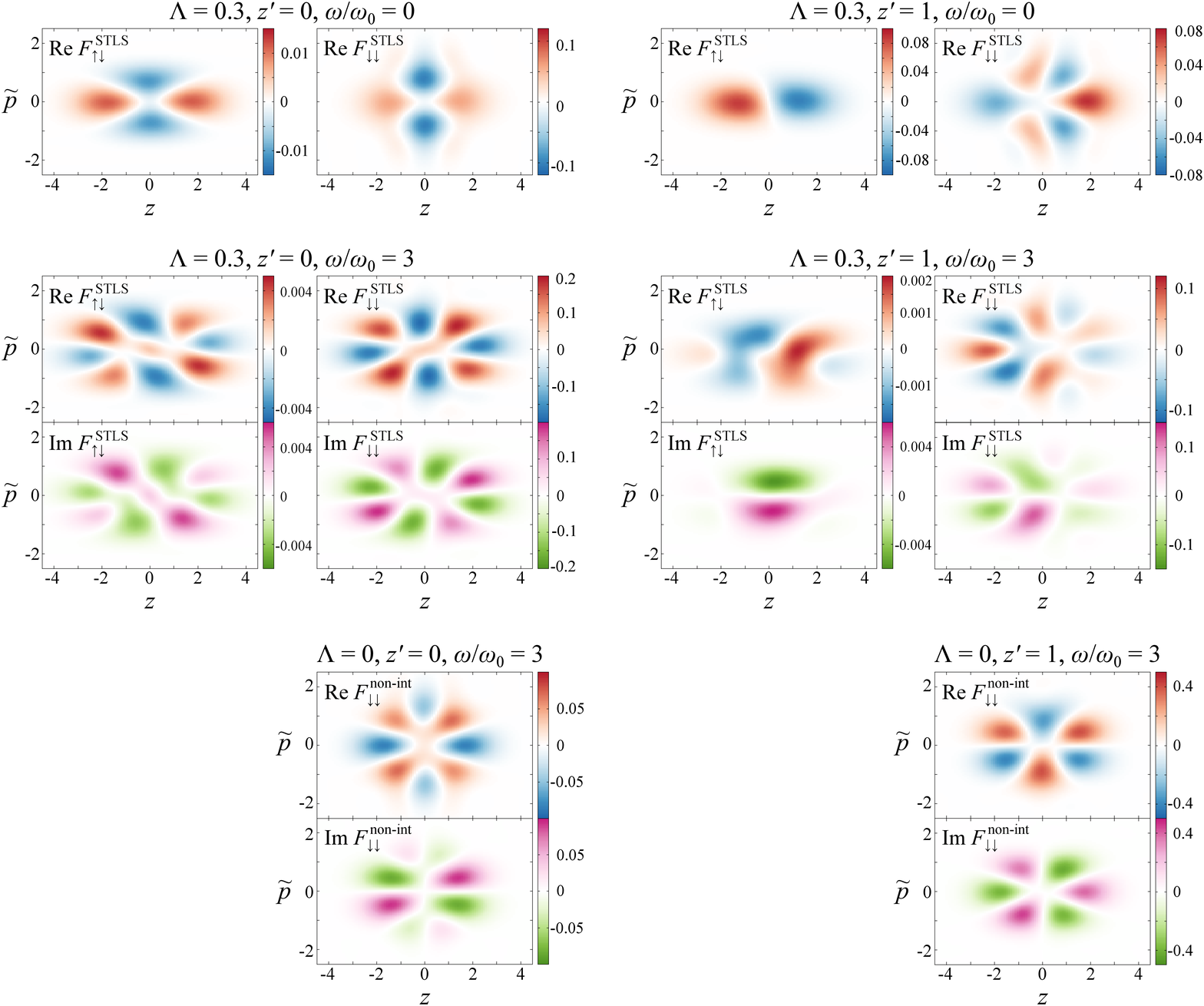}
\label{Fig_harm_resp_WDF1}
\end{figure*}

In Fig. \ref{Fig_harm_resp_WDF1},
the STLS-approximated $F^{\mathrm{STLS}}$ and the non-interacting $F^{\mathrm{non-int}}$ scalar response of
the one-particle WDF are shown for spin-down perturbation applied at $z' = 0$ and $1$.
We solved the matrix equation for the largest order $n_{\mathrm{max}}^{\mathrm{EOM}} = 20$ to calculate the expansion coefficients,
from which we adopted those up to the order $n_{\mathrm{max}}^{\mathrm{resp}} = 10$ to calculate $F^{\mathrm{STLS}}$.
We found that $n_{\mathrm{max}}^{\mathrm{EOM}}$ must be larger than $n_{\mathrm{max}}^{\mathrm{resp}}$ for avoiding the effects of a truncated matrix which is originally infinite dimensional. 
The highest order of truncation for $F^{\mathrm{STLS}}$ was confirmed to be sufficiently large to capture their overall behavior for the parameters used here.
$F_{\uparrow \downarrow}^{\mathrm{non-int}}$ vanishes since only the collision term affects
the dynamics of electrons having the spin direction opposite to perturbation [see eq. (\ref{def_collision_func})].
It is observed in the figure that, for given $\Lambda, z'$ and $\omega$, the shape of $F$ as a function of $z$ and $\widetilde{p}$ is more nodal for the larger $\omega$. 
This result is natural since $F$ for the larger $\omega$ should be responsible for the higher-energy excitation contributing to the density response function.

\subsubsection{Density response function}

\begin{figure*}[p]
\caption{
For the interacting two-electron system,
the exact $\chi$ (solid curves) and STLS-approximated $\chi^{\mathrm{STLS}}$ (dashed curves) as functions of $\omega$
are plotted for various combinations of $z, z'$, and $\Lambda$.
$m$ is set to unity for the plot.
}
\includegraphics[keepaspectratio,width=16cm]{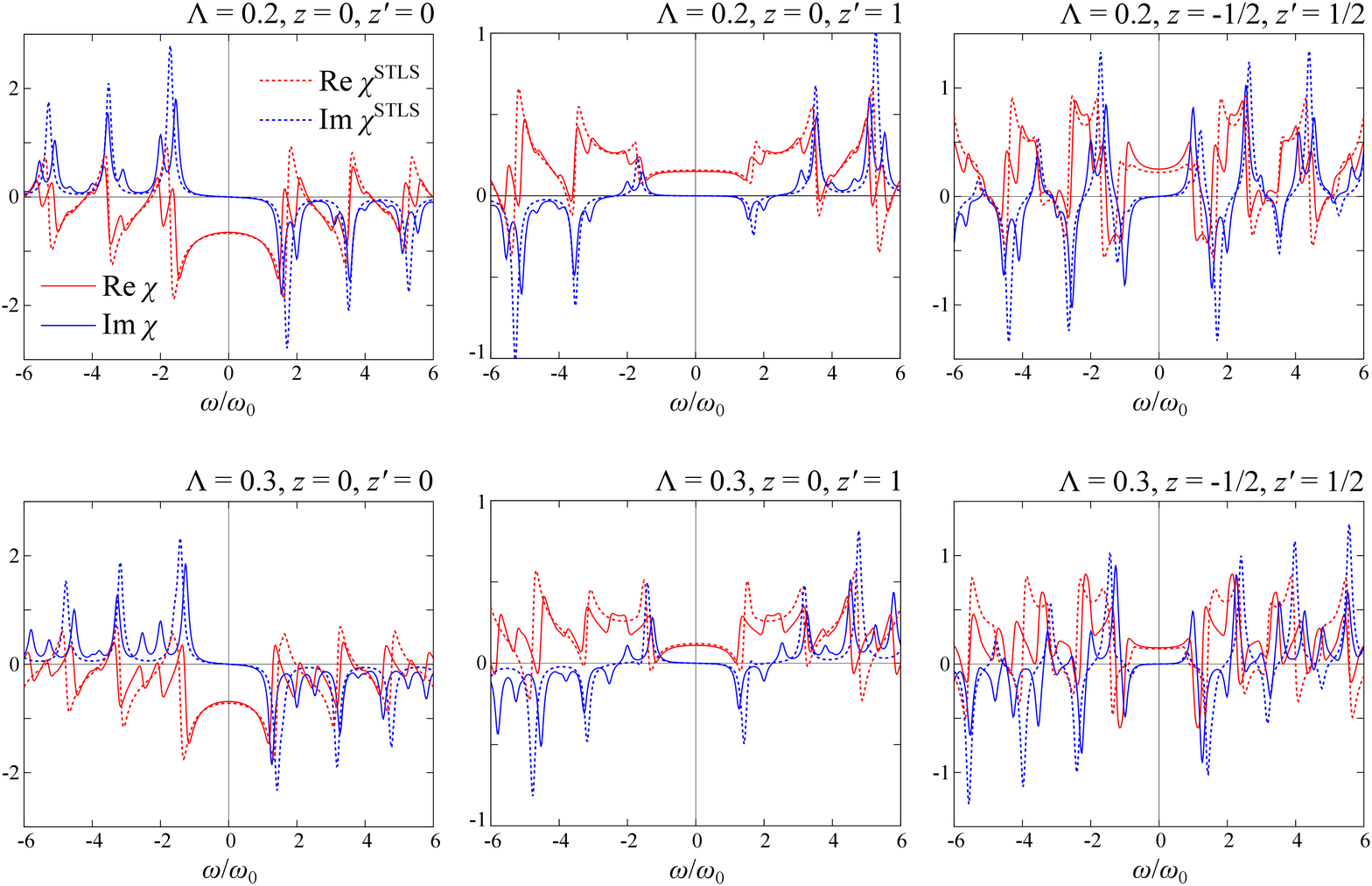}
\label{Fig_chi_harm}
\end{figure*}

Substituting the coefficients obtained by solving eq. (\ref{harm_harm_resp_coeff_recurr_exact_twoel}) into eq. (\ref{harm_harm_resp_in_hermite})
and integrating over the momentum,
we get the spin-dependent density response function [see eq. (\ref{resp_of_el_dens_to_W})]
\begin{gather}
	\chi_{\sigma \sigma'}^{\mathrm{STLS}} (x, x', \omega)
	=
		\frac{2 m e^{-z^2/2}}{\sqrt{\pi}}
		\sum_{n}
			c_{\sigma n 0} (\sigma', x', \omega)
			H_n \left( \frac{z}{\sqrt{2}} \right)
	.
	\label{harm_harm_chi_for_W_ext_in_Hermite}
\end{gather}
Since we have used the exact $g_{\sigma \sigma'}$ and $f_{\sigma}^{(0)}$ in the present case,
the approximation introduced to obtain $\chi^{\mathrm{STLS}}$ is only the STLS approximation.
It is thus possible to examine the validity of STLS approximation by comparing $\chi^{\mathrm{STLS}}$ with the exact one.
For various combinations of the dimensionless positions $z, z'$, and the interaction strength $\Lambda$,
$\chi^{\mathrm{STLS}}$ and the exact response function $\chi$  are plotted in Fig. \ref{Fig_chi_harm}.
We solved the matrix equation for $n_{\mathrm{max}}^{\mathrm{EOM}} = 20$ to calculate the expansion coefficients,
from which we adopted those up to the order $n_{\mathrm{max}}^{\mathrm{resp}} = 16$ to calculate the response from eq. (\ref{harm_harm_resp_in_hermite}).
For the calculation of $\chi$, we incorporated only the contributions for $n_{\mathrm{c}} + n_{\mathrm{r}} \leqq 16$ in eq. (\ref{harm_harm_chi_even_nc_nr}).
The highest orders of truncation for $\chi$ and $\chi^{\mathrm{STLS}}$ were confirmed to be sufficiently large to capture their overall behavior for the range of $\omega$ in the figure.
It is seen that $\chi$ and $\chi^{\mathrm{STLS}}$ are in good agreement for $|\omega/\omega_0| \leqq 1$,
while the discrepancy between them begins to appear as $|\omega|$ becomes larger.
For given $z$ and $z'$, the deviation of $\chi^{\mathrm{STLS}}$ from $\chi$ for $\Lambda = 0.3$ as a function of $\omega$
shows a tendency to be larger than that for $\Lambda = 0.2$,
implying that the STLS approximation is worse for the stronger interaction. 
The most striking differences between the exact and the approximated response functions are seen in their peak structures of imaginary parts,
which represent the electron-number-conserving excitations.
The weak peaks near a strong one in the interacting $\chi$ are degenerate for the non-interacting case.
Such subpeaks are not reproduced by $\chi^{\mathrm{STLS}}$.
As is seen, $\chi^{\mathrm{STLS}}$ exhibits only the smeared features of $\chi$
and the significant peaks which tend to be stronger than those found in $\chi$.
There exists another tendency that a peak in $\chi$ undergoes blue shift in $\chi^{\mathrm{STLS}}$ if it is found there. 
It is of interest whether these tendencies of STLS approximation are generic or not.
Comparison of the exact and STLS-approximated response functions for other interacting systems will help us to understand the behavior of inhomogeneous STLS approach.

\subsubsection{Interaction energy}

\begin{figure}
\includegraphics[keepaspectratio,width=7cm]{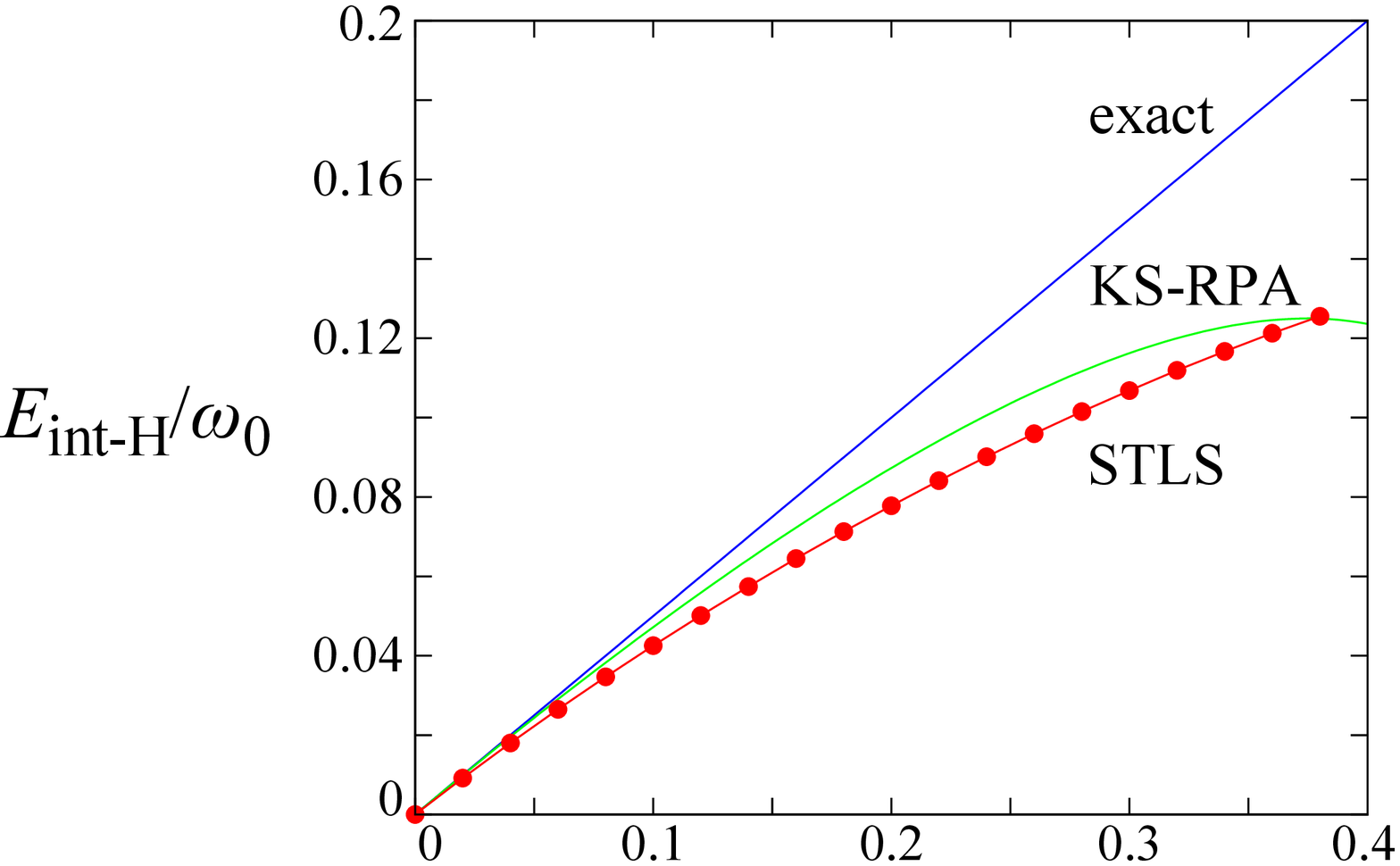}
\caption{
For the interacting two-electron system,
the exact $E_{\mathrm{int-H}}$,
the STLS-approximated $E_{\mathrm{int-H}}^{\mathrm{STLS}}$, and
the Kohn-Sham $E_{\mathrm{int-H}}^{\mathrm{KS-RPA}}$
differences between the interaction energy and the Hartree energy are plotted as functions of the interaction strength $\Lambda$.
}
\label{Fig_e_int-H}
\end{figure}

The interaction energy for the ground state is calculated from the exact two-electron distribution in eq. (\ref{harm_harm_two_el_gs}) as
$E_{\mathrm{int}} = -\omega_0 (1 - \lambda^2)/(4 \lambda)$,
while the Hartree energy is calculated from eq. (\ref{harm_harm_el_dens_gs}) as
$E_{\mathrm{H}} = -\omega_0 (1 - \lambda) (1 + \lambda)^2 /(4 \lambda)$.
The difference between them is thus
\begin{gather}
	E_{\mathrm{int-H}}
	=
		\omega_0
		\frac{1 - \lambda^2}{4}
	=
		\omega_0
		\frac{\Lambda}{2}
	.
\end{gather}
By plugging the expression of STLS response function in eq. (\ref{harm_harm_chi_for_W_ext_in_Hermite}) into eq. (\ref{E_corr_FDT}),
we can write the STLS-approximated value of $E_{\mathrm{int-H}}$ as
\begin{widetext}
\begin{gather}
	E_{\mathrm{int-H}}^{\mathrm{STLS}}
	=
		\omega_0
		\frac{\Lambda}{8 \pi}
		\int_{-\infty}^\infty
		d z' 
		\int_0^\infty
		\frac{d \omega}{\omega_0}
			\sum_{\sigma, \sigma'}
			\mathrm{Im}
			\left[
				4 \sqrt{2}
				c_{\sigma 2 0} (\sigma', x', \omega)
				-
				4 
				z'
				c_{\sigma 1 0} (\sigma', x', \omega)
				+
				(z'^2 + 1)
				\sqrt{2}
				c_{\sigma 0 0} (\sigma', x', \omega)
			\right]
	.
\end{gather}
\end{widetext}
In Fig. \ref{Fig_e_int-H},
$E_{\mathrm{int-H}}$ and $E_{\mathrm{int-H}}^{\mathrm{STLS}}$ are plotted as functions of $\Lambda$.
It is seen that the discrepancy between the exact and the STLS values becomes larger as $\Lambda$ becomes larger,
showing the slightly stronger variation than linear in $\Lambda$.
These results are consistent with the tendency of error in $f_{\sigma \sigma'}^{(0) \mathrm{STLS}}$ discussed above.

It is interesting to calculate the interaction energy by adopting the Kohn-Sham response function as often done in DFT calculations.
Since the correlation effects are effectively incorporated in the non-interacting KS system
and we know the exact  $\chi_{\sigma \sigma'}^{\mathrm{KS}}$ together with its expansion coefficients in the present case (see Appendix),
we can calculate the KS-RPA value analytically as
\begin{gather}
	E_{\mathrm{int-H}}^{\mathrm{KS-RPA}}
	=
		\omega_0
		\Lambda
		\frac{\lambda}{1 + \lambda}
	,
\end{gather}
plotted in Fig. \ref{Fig_e_int-H}.
It is seen that the agreement between the exact and the KS-RPA values for $\Lambda < 0.38$ is better than that between the exact and the STLS values. 
In the stronger-interaction regime, the STLS values appear to be better.
The reliability of KS-RPA and STLS interaction energies may be system dependent 
and further examinations will be needed for various systems.

\section{Conclusions}

By adopting the usual STLS approximation, which factorizes the two-particle WDF into the two one-particle WDFs and the static pair correlation function,
we proposed a quantum STLS approach for inhomogeneous interacting electronic systems under a time-dependent electromagnetic perturbation.
Specifically, we started from the second-quantized Hamiltonian and
derived the linear EOM for the response of the one-particle WDF to the perturbation via the functional derivatives.
It was demonstrated that the EOM under the slow-variation approximation for the inherent potential and the electronic interaction
becomes a one having the similar form adopted by the semi-classical inhomogeneous STLS approach.\cite{bib:4075, bib:4076}

Since there exist multiple sources of errors in a self-consistent inhomogeneous STLS calculation,
we chose an interacting confined two-electron system to extract the error coming only from the STLS approximation.
We were able to calculate the exact response function and the WDFs from the analytic solutions,
with which we compared the STLS response function.
It was analytically demonstrated for the system that the STLS approximation for equilibrium WDFs become worse as the interaction gets stronger.
While the STLS response function $\chi^{\mathrm{STLS}}$ reproduced well the exact one $\chi$ for low-energy excitations,
the discrepancy between them becomes larger for the higher-energy excitations.
The STLS approximation was found to tend to reproduce the strong peaks in the imaginary part of $\chi$ with blue shift and
fail to reproduce the weak exact peaks which are merged to their nearby strong peaks in the non-interacting case.
We calculated numerically the contribution to interaction energy from $\chi^{\mathrm{STLS}}$
to demonstrate that the discrepancy between the exact and the STLS values are larger for the larger $\Lambda$, as expected.   
While we focused on the comparison between $\chi$ and $\chi^{\mathrm{STLS}}$ since the latter is directly related to the formulation of inhomogeneous STLS approach,
that between the exact $f_{\mathrm{xc}}$ and the STLS $f_{\mathrm{xc}}^{\mathrm{STLS}}$ exchange correlation kernels will also provide useful insights if they can be constructed.\cite{bib:3998}

Further examinations on the validity of STLS approximation are needed for various inhomogeneous interacting systems whose WDFs are ideally analytically calculated.\cite{bib:3946}
The quality of STLS approximation when applied to periodic systems will be, regardless of achieving self-consistency,
interesting particularly for electronic-structure calculations of solids.
One of the reasons from the viewpoint of computational resources for the successful applications of the semi-classical inhomogeneous STLS approach formulated by Dobson et. al.\cite{bib:4075}
is the derivation of integral equations for response functions where the momentum variables are absent,
in contrast to the EOM derived in the present work.
If one wants to calculate only the STLS correlation energy for an interacting system,
there is no need to know the momentum dependence of the response of one-particle WDF
since only its integral over momentum is required in FDT.
That is often the case for electronic-structure calculations of realistic systems.
If a method in which the momentum variables have been integrated out from the EOM is constructed,
it will make the quantum inhomogeneous STLS approach more tractable for realistic applications. 
From the viewpoint of quantum chemistry,
a molecular system for which the full-configuration interaction (FCI) calculation is possible can be used for the examination of the validity of our STLS approach
since the exact interacting one- and two-particle WDFs are extractable from the FCI results in principle.

\acknowledgments

This research was supported by MEXT as ”Exploratory Challenge on Post-K computer” (Frontiers of Basic Science: Challenging the Limits). This research used computational resources of the K computer provided by the RIKEN Advanced Institute for Computational Science through the HPCI System Research project (Project ID:hp160265).

\appendix

\section{Derivation of EOM for a one-particle WDF}

With the slow-variation approximation to the vector potential in eq. (\ref{slow_var_approx_for_vec_pot}),
the contribution from the kinetic-energy operator to eq. (\ref{EOM_Winger_opr_H}) is calculated as
\begin{gather}
	[ \hat{f}_{\sigma} (\boldsymbol{r}, \boldsymbol{p}, t), \hat{T} ]
	=
		-
		\frac{i}{m}
		\left[
			\boldsymbol{p}
			+
			\frac{1}{c}
			\boldsymbol{A}_{\sigma} (\boldsymbol{r}, t)
		\right]
		\cdot
		\nabla_{\boldsymbol{r}}
		\hat{f}_\sigma (\boldsymbol{r}, \boldsymbol{p}, t)
	\nonumber \\
		+
		\frac{i}{m c}
		\sum_j
			\left[ \boldsymbol{p} + \frac{1}{c} \boldsymbol{A}_{\sigma} ( \boldsymbol{r}, t) \right]
			\cdot
			\frac{\partial \boldsymbol{A}_{\sigma} (\boldsymbol{r}, t)}{\partial r_j}
			\frac{\partial \hat{f}_\sigma (\boldsymbol{r}, \boldsymbol{p}, t)}{\partial p_j}		
	.
	\label{WDF1_EOM_kin_slow_vec_pot}
\end{gather}

The chain rule for the variable transformation introduced in (\ref{variable_transform}) leads to the following relations among the derivatives with respect to the old and new variables: 
\begin{gather}
	\frac{\partial}{\partial r_j}
	=
		\frac{\partial}{\partial \widetilde{r}_j}
		+
		\frac{1}{c}
		\frac{\partial \boldsymbol{A}_{\sigma} (\boldsymbol{r}, t)}{\partial r_j}
		\cdot
		\frac{\partial}{\partial \boldsymbol{\Pi}}
	\label{deriv_new_r}
	\\
	\frac{\partial}{\partial p_j}
	=
		\frac{\partial}{\partial \Pi_j}
	\label{deriv_pi}
	\\
	\frac{\partial}{\partial t}
	=
		\frac{1}{c}
		\frac{\partial \boldsymbol{A}_{\sigma} (\boldsymbol{r}, t)}{\partial t}
		\cdot
		\frac{\partial}{\partial \boldsymbol{\Pi}}
		+
		\frac{\partial}{\partial \widetilde{t}}
	\label{deriv_new_t}
\end{gather}
for $j = x, y, z$.
The Jacobian for this transformation is easily confirmed to be unity.
We rewrite eq. (\ref{WDF1_EOM_kin_slow_vec_pot}) expressed in the old variables
by using eqs. (\ref{deriv_new_r}) - (\ref{deriv_new_t}) and the replacement in eq. (\ref{replace_p_by_pi}) as
\begin{gather}
	[ \hat{\widetilde{f}}_{\sigma} (\widetilde{\boldsymbol{r}}, \boldsymbol{\Pi}, \widetilde{t}), \hat{T} ]
	=
		-\frac{i}{m}
		\boldsymbol{\Pi}
		\cdot
		\nabla_{\widetilde{\boldsymbol{r}}}
		\hat{\widetilde{f}}_\sigma
		+
		\frac{i}{m c}	
		(\boldsymbol{\Pi} \times \boldsymbol{B}_{\sigma})
		\cdot
		\nabla_{\boldsymbol{\Pi}}
		\hat{\widetilde{f}}_\sigma
	,
	\label{WDF_EOM_kin_slow_vec_pot_B}
\end{gather}
where we have used the relation
$
\boldsymbol{B}_{\sigma} (\boldsymbol{r}, t)
=
\nabla_{\boldsymbol{r}} \times \boldsymbol{A}_{\sigma} (\boldsymbol{r}, t)
$
between the magnetic field and the vector potential.
The second term on the right-hand side in eq. (\ref{WDF_EOM_kin_slow_vec_pot_B}) represents the Lorentz force exerted  on an electron.

By using eq. (\ref{deriv_new_t})
and the fact that the function form of the vector potential is invariant under the transformation, that is,
$\boldsymbol{A}_{\sigma} (\boldsymbol{r}, t) = \widetilde{\boldsymbol{A}}_{\sigma} (\widetilde{\boldsymbol{r}}, \widetilde{t})$,
the left-hand side of eq. (\ref{EOM_Winger_opr_H}) is rewritten as
\begin{gather}
	i
	\frac{\partial \hat{f}_{\sigma}}{\partial t} 
	=
		i
		\frac{\partial \hat{\widetilde{f}}_\sigma}{\partial \widetilde{t}} 
		+
		\frac{i}{c}
		\frac{\partial \boldsymbol{A}_{\sigma}}{\partial \widetilde{t}}
		\cdot
		\nabla_{\boldsymbol{\Pi}} \hat{\widetilde{f}}_{\sigma}
	.
	\label{WDF1_lhs_vec_pot}
\end{gather}

For an arbitrary function $C (\boldsymbol{r}, \boldsymbol{r}')$,
we define a shorthand notation
$
	C^{\mathrm{(W)}} (  \boldsymbol{r}, \boldsymbol{r}' )
	\equiv
			C (  \boldsymbol{r} + \boldsymbol{r}' )
			-
			C (  \boldsymbol{r} - \boldsymbol{r}' )
	.
$
With this,
the contribution from the potential part in the Hamiltonian to the right-hand side of eq. (\ref{EOM_Winger_opr_H}) is written as
\begin{gather}
	[ \hat{\widetilde{f}}_{\sigma} (\widetilde{\boldsymbol{r}}, \boldsymbol{\Pi}, \widetilde{t}), \hat{V} ]
	=
	\nonumber \\
		-
		\int
			\frac{d^3 \overline{r}}{h^3}
			d^3 \Pi' \,
			e^{i ( \boldsymbol{\Pi} - \boldsymbol{\Pi}' ) \cdot \overline{\boldsymbol{r}}}
			V_{\sigma}^{\mathrm{(W)}} \left(  \widetilde{\boldsymbol{r}} , \frac{\overline{\boldsymbol{r}}}{2} , \widetilde{t} \right)
			\hat{\widetilde{f}}_{\sigma} (\widetilde{\boldsymbol{r}}, \boldsymbol{\Pi}', \widetilde{t})
	.
	\label{WDF1_rhs_V_vec_pot}
\end{gather}
The contribution from the interaction part to the right-hand side of eq. (\ref{EOM_Winger_opr_H}) is written similarly as
\begin{gather}
	[ \hat{\widetilde{f}}_{\sigma} (\widetilde{\boldsymbol{r}}, \boldsymbol{\Pi}, \widetilde{t}), \hat{H}_{\mathrm{int}} ]
	=
		-
		\int
		\frac{d^3 \overline{r}}{h^3}
		d^3 r'
		d^3 \Pi_1
		d^3 \Pi_2 \,
			e^{i ( \boldsymbol{\Pi} - \boldsymbol{\Pi}_1  ) \cdot \overline{\boldsymbol{r}}}
	\cdot
	\nonumber \\
	\cdot	
			v^{\mathrm{(W)}} \left( \widetilde{\boldsymbol{r}}  - \boldsymbol{r}', \frac{\overline{\boldsymbol{r}}}{2} \right)
			\sum_{\sigma'}
			\hat{\widetilde{f}}_{\sigma \sigma'} (\widetilde{\boldsymbol{r}}, \boldsymbol{\Pi}_1, \boldsymbol{r}', \boldsymbol{\Pi}_2, \widetilde{t})
	,
	\label{WDF1_rhs_H_int_vec_pot}
\end{gather}
where $\hat{\widetilde{f}}_{\sigma \sigma'}$ is the two-particle Wigner distribution operator in the transformed variables defined similarly to the one-particle operator.

Collecting eqs. (\ref{WDF_EOM_kin_slow_vec_pot_B}), (\ref{WDF1_lhs_vec_pot}), (\ref{WDF1_rhs_V_vec_pot}), and (\ref{WDF1_rhs_H_int_vec_pot})
for eq. (\ref{EOM_Winger_opr_H}) and taking the expectation values of both sides,
we obtain the EOM for the one-particle WDF in eq. (\ref{EOM_WDF1_with_vec_pot}).

\section{WDFs for a two-electron system}

For an interacting electronic system in equilibrium at a zero temperature,
the expectation values of operators in the integrands in eqs. (\ref{def_Wigner_opr_collinear}) and (\ref{def_Wigner_opr_collinear_2}) are
expressed in the many-electron wave function $\Psi_0$ of the ground state.\cite{bib:Stefanucci_and_Leeuwen}
In particular for a two-electron system,
if it is factorized into the spatial $\psi (\boldsymbol{r}_1, \boldsymbol{r}_2) $ and spin $\phi (\sigma_1, \sigma_2)$ parts as
$\Psi_0 (\boldsymbol{r}_1, \sigma_1, \boldsymbol{r}_2, \sigma_2) = \psi (\boldsymbol{r}_1, \boldsymbol{r}_2) \phi (\sigma_1, \sigma_2)$,
the one-particle WDF is calculated as
\begin{widetext}
\begin{gather}
	f_{\sigma}^{(0)} (\boldsymbol{r}, \boldsymbol{p})
	=
		\frac{1}{h^3}
		\int
		d^3 \overline{r}
		d^3 r' \,
			e^{i \boldsymbol{p} \cdot \overline{\boldsymbol{r}}}
			\psi \left( \boldsymbol{r}', \boldsymbol{r} + \frac{\overline{\boldsymbol{r}}}{2} \right)^*
			\psi \left( \boldsymbol{r}', \boldsymbol{r} - \frac{\overline{\boldsymbol{r}}}{2} \right)
			\sum_{\sigma'}
				| \phi (\sigma', \sigma) |^2
	\label{WDF1_from_two_el_Psi}
\end{gather}
and the two-particle WDF is calculated as
\begin{gather}
	f_{\sigma_1 \sigma_2}^{(0)} (\boldsymbol{r}_1, \boldsymbol{p}_1, \boldsymbol{r}_2, \boldsymbol{p}_2)
	=
		\frac{1}{h^6}
		\int
		d^3 \overline{r}_1
		d^3 \overline{r}_2
			e^{i \boldsymbol{p}_1 \cdot \overline{\boldsymbol{r}}_1 + i \boldsymbol{p}_2 \cdot \overline{\boldsymbol{r}}_2}
			\psi \left( \boldsymbol{r}_2 + \frac{\overline{\boldsymbol{r}}_2 }{2}, \boldsymbol{r}_1 + \frac{\overline{\boldsymbol{r}}_1 }{2} \right)^*
			\psi \left( \boldsymbol{r}_2 - \frac{\overline{\boldsymbol{r}}_2 }{2},  \boldsymbol{r}_1 - \frac{\overline{\boldsymbol{r}}_1 }{2} \right)
			| \phi (\sigma_2, \sigma_1)|^2
	.
	\label{WDF2_from_two_el_Psi}
\end{gather}
\end{widetext}

\section{Density response function for a two-electron system}

The expression of density response function for an interacting electronic system is 
given often in the well known Lehmann representation.\cite{bib:Stefanucci_and_Leeuwen}
In particular for a two-electron system whose energy eigenfunctions $\Psi_{\nu}$ are all factorizable as above,
the spin-dependent density response function at a zero temperature is calculated as a summation over all the excited states:
\begin{gather}
	\chi_{\sigma_1 \sigma_2} (\boldsymbol{r}_1, \boldsymbol{r}_2, \omega)
	=
		\sum_{{\nu \ne 0}}
			\Bigg[
				\frac{
					P_{0 \nu} ( \boldsymbol{r}_1 )
					P_{0 \nu} ( \boldsymbol{r}_2 )^*
					S_{0 \nu \sigma_1}
					S_{0 \nu \sigma_2}^*
				}{\omega - ( E_\nu - E_0 ) + i \delta}
	\nonumber \\
				-
				\frac{
					P_{0 \nu} ( \boldsymbol{r}_1 )^*
					P_{0 \nu} ( \boldsymbol{r}_2 )
					S_{0 \nu \sigma_1}^*
					S_{0 \nu \sigma_2}
				}{\omega + ( E_\nu - E_0 ) + i \delta}
			\Bigg]
	,
	\label{dens_reps_func_two_el}
\end{gather}
where the transition amplitudes for the spatial and spin parts are given, respectively, by
\begin{gather}
	P_{\nu \nu'} (\boldsymbol{r})
	\equiv
		\int
		d^3 r' \,
		\psi_{\nu} (\boldsymbol{r}', \boldsymbol{r} )^*	
		\psi_{\nu'} (\boldsymbol{r}', \boldsymbol{r} )
	,
	\label{def_P_for_dens_resp_two_el}
	\\
	S_{\nu \nu' \sigma}
	\equiv
		\sum_{\sigma'}
			\phi_{\nu} (\sigma', \sigma)^*
			\phi_{\nu'} (\sigma', \sigma)
	.
	\label{def_S_for_dens_resp_two_el}
\end{gather}
$\delta$ is a positive infinitesimal constant for ensuring causality.

In particular for a non-interacting two-electron system,
the expression of the density response function is given in terms of one-particle orbitals as\cite{bib:Stefanucci_and_Leeuwen}
\begin{gather}
	\chi_{\sigma \sigma'}^{\mathrm{non-int}} (\boldsymbol{r}_1, \boldsymbol{r}_2, \omega)
	=
		\delta_{\sigma \sigma'}
		\sum_{n \ne 0}
			\Bigg[
				\frac{\psi_0 (\boldsymbol{r}_1) \psi_n (\boldsymbol{r}_1) \psi_0 (\boldsymbol{r}_2)^*  \psi_n (\boldsymbol{r}_2)^*}{\omega - (\varepsilon_n - \varepsilon_0 )+ i \delta}
	\nonumber \\
				-
				\frac{\psi_0 (\boldsymbol{r}_1)^* \psi_n (\boldsymbol{r}_1)^* \psi_0 (\boldsymbol{r}_2)  \psi_n (\boldsymbol{r}_2)}{\omega + (\varepsilon_n - \varepsilon_0) + i \delta}
			\Bigg]
	,
	\label{harm_harm_dens_resp_non_int_as_sum}
\end{gather}
where $\psi_n$ is the spatial part of the $n$-th one-particle orbital with its orbital energy $\varepsilon_n$.
The factor $2$ is the spin degeneracy.

\section{Integral formula}

It is known that the Hermite polynomials satisfy the recursion relation
\begin{gather}
 	\frac{d}{d x} [ e^{-x^2} H_n (x) ]
	=
		- e^{-x^2}
		H_{n + 1} (x)
	.
	\label{deriv_Hermite}
\end{gather}
For an $a > 0$,
we use this relation repeatedly via partial integration to get
\begin{gather}
	I_{n_1 n_ 2} (y ; a)
	\equiv
	\nonumber \\
		\int_{-\infty}^\infty
		d x \,
			e^{- (x + y)^2 - a^2 (x - y)^2}
			H_{n_1} (x + y)
			H_{n_2} (a (x - y) )
	\nonumber \\
	=
		-a
		I_{n_1 - 1, n_2 + 1} (y; a)
	=
	\cdots
	=
		(- a)^{n_1}
		I_{0, n_1 + n_2} (y; a)	
	\nonumber \\
	=			
		(-1)^{n_2}
		a^{n_1}
		\exp
		\left( -\frac{4 a^2}{1 + a^2} y^2 \right)
	\cdot
	\nonumber \\
	\cdot
		\frac{\sqrt{\pi}}{(1 + a^2)^{(n_1 + n_2 + 1)/2}}
		H_{n_1 + n_2} \left( \frac{2 a}{\sqrt{1 + a^2}} y \right)
	,
	\label{integ_exp_exp_H_H}
\end{gather}
where we used an integral formula\cite{bib:Gradshteyn_and_Ryzhik}
\begin{gather}
	\int_{-\infty}^\infty
	d x \,
		e^{- (x - y)^2/ (2 u)}
		H_n (x)
	\nonumber \\
	=
		\sqrt{2 \pi u}
		(1 - 2 u)^{n/2}
		H_n \left( \frac{y}{\sqrt{1 - 2 u}} \right)
	\label{integ_displaced_Gauss_H}
\end{gather}
for a $u \geqq 0$ to reach the last equality in eq. (\ref{integ_exp_exp_H_H}).

\section{Integrals appearing in EOM for the two-electron system}

The integrals involving the Hermite polynomials below can be evaluated analytically by using the formulae in the literature.\cite{{bib:Gradshteyn_and_Ryzhik}}

Those in the first summation on the left-hand side in eq. (\ref{harm_harm_resp_coeff_recurr_exact_twoel}) are given by
\begin{widetext}
\begin{gather}
	I_{n}^{(1)}
	\equiv
		\frac{1}{2^n n! \sqrt{\pi}}
		\int_{-\infty}^\infty
		d z \,
			z
			\exp \left( -\frac{4}{1 + \lambda} z^2 \right)	
			H_n (z)
	=
		\begin{cases}
			0 & \mathrm{for} \, \mathrm{even} \, n \\
			\frac{1}{2^{n} n! \sqrt{\pi}}
			(\lambda - 3)^{(n - 1)/2}
			(1 + \lambda)^{3/2}
			\Gamma (n/2 + 1) /4
			& \mathrm{for} \, \mathrm{odd} \, n
		\end{cases}
	.
\end{gather}
and
\begin{gather}
	M_{n n_1}
	\equiv
		\frac{\sqrt{2}}{(1 + \lambda) 2^{n - 1} n! \pi}
		\int_{-\infty}^\infty
		d z \,
			H_n \left( \frac{z}{\sqrt{2}} \right)
			\exp \left[ - \frac{(1 + \lambda)^2}{2 (3 + \lambda^2)} z^2 \right]
			[ M_{n_1}^{(1)} (z)  - M_{n_1}^{(2)} (z) ]
	,
	\label{def_integ_M_1}
\end{gather}
\end{widetext}
where
\begin{gather}
	M_{n_1}^{(1)} (z)
	\equiv
		z
		\int_{-\infty}^\infty
		d z' \,
			M (z, z')
			H_{n_1} \left( \frac{z'}{\sqrt{2}} \right)
\end{gather}
and
\begin{gather}
	M_{n_1}^{(2)} (z)
	\equiv
		\int_{-\infty}^\infty
		d z' \,
			M (z, z')
			z'
			H_{n_1} \left( \frac{z'}{\sqrt{2}} \right)
	\label{harm_harm_integ_M2}
\end{gather}
for
\begin{gather}
	M (z, z')
	\equiv
		\exp
		\left[
			- \frac{3 + \lambda^2}{4 (1 + \lambda)}
			\left( \frac{1 - \lambda^2}{3 + \lambda^2} z + z' \right)^2
		\right]
	.
\end{gather}
That in the second summation is given by
\begin{gather}
	I_{n n'}^{(2)}
	\equiv
		\frac{1}{2^n n! \sqrt{\pi}}
		\int_{-\infty}^\infty
		d z \,
			z
			e^{- z^2}
			H_{n} (z)
			H_{n'} (z)	
	\nonumber \\
	=
		\begin{cases}
			n + 1 & \mathrm{for} \, n' = n + 1 \\
			\frac{1}{2} & \mathrm{for} \, n' = n - 1 \\
			0 & \mathrm{otherwise}
		\end{cases}
		.
\end{gather}
The integration in eq. (\ref{def_integ_M_1}) can be performed analytically as well as $I_n^{(1)}$ and $I_{n n'}^{(2)}$.

The integral on the right-hand side in eq. (\ref{harm_harm_resp_coeff_recurr_exact_twoel}) is given by
\begin{widetext}
\begin{gather}
	B_{n n'} (z') 
	\equiv
		\sqrt{\frac{\lambda}{1 + \lambda}}
		\frac{1}{2^{n + n' - 1} n! n'! \pi^{3/2}}
		\int_{-\infty}^\infty
		d \widetilde{p} \,
			H_{n'} (\widetilde{p})
			\mathrm{Im} \,  \widetilde{B}_n ( z', \widetilde{p})
	,
\end{gather}
where
\begin{gather}
	\widetilde{B}_n ( z', \widetilde{p})
	\equiv
		\frac{1}{\sqrt{2}}
		\int_{-\infty}^\infty
		d z \,
			H_n \left( \frac{z}{\sqrt{2}} \right)
			e^{ i 2 \widetilde{p} (z' - z) }
			\exp
			\left[
				- \frac{1 + \lambda}{4} z'^2
				+ \frac{1 + \lambda}{2} z z'
				- \frac{\lambda^2 + 6 \lambda + 1}{4 (1 + \lambda)} z^2
			\right]
	\nonumber \\
	=
		\exp
		\Bigg[
				-2 u
				\Bigg(
					\widetilde{p}
					-
					\frac{i [1 - u (1 + \lambda)/2 ]}{2 u}
					z'
				\Bigg)^2
				-
				\frac{ 1 - u (1 + \lambda)/2  }{2 u} z'^2
		\Bigg]
		\sqrt{\pi u}
		(1 - u)^{n/2}
		H_n
		\left(
			\frac{  (u/\sqrt{2}) [ (1 + \lambda) z'/2 - i 2 \widetilde{p} ] }{ \sqrt{1 - u} }
		\right)
	\label{harm_harm_def_coeff_pert}
\end{gather}
\end{widetext}
is defined by using the formula in eq. (\ref{integ_displaced_Gauss_H}) for $u \equiv 2 (1 + \lambda)/(\lambda^2 + 6 \lambda + 1)$.
$B_{n n'} (z') $ can be nonzero only when $n'$ is odd since the imaginary part of $\widetilde{B}_n ( z', \widetilde{p})$ is odd with respect to $\widetilde{p}$.

\section{Another form of EOM for the two-electron system in non-interacting case}

In the non-interacting case, the expansion coefficients $c_{\sigma n n'} (\sigma', x', \omega)$ in eq. (\ref{harm_harm_resp_in_hermite})
can be nonzero only for $\sigma = \sigma'$ and is independent of the spin direction.
By redefining the coefficients in eq. (\ref{harm_harm_resp_coeff_recurr_exact_twoel}) as $c_n^{(s)} \equiv c_{\sigma n, s - n}$,
we rewrite the linear equation as
\begin{gather}
	i
	\frac{\omega}{\omega_0}
	c_n^{(s)}
	+
	\frac{\sqrt{2}}{4}
	c_{n - 1}^{(s - 2)}
	+
	\sqrt{2}
	(s - n + 1)
	c_{n - 1}^{(s)}
	-
	\frac{\sqrt{2}}{2}
	(n + 1)
	c_{n + 1}^{(s)}
	\nonumber \\
	=
		B_{n, s - n} (z')
	.
\end{gather}
For a given $s$, one can get $c_n^{(s)} (n = 0, 1, \dots, s)$ by solving this $(s + 1)$-dimensional linear equation
if $c_n^{(s - 2)} (n = 0, 1, \dots, s - 2)$ has been known.
The exact non-interacting response $F^{\mathrm{non-int}}$ for this case can thus be calculated without being worried about the truncation errors.

\section{Non-interacting response function for the two-electron system}

In the non-interacting case, the expansion coefficients $c_{\sigma n 0} (\sigma', x', \omega)$ in eq. (\ref{harm_harm_chi_for_W_ext_in_Hermite}) can be calculated directly.
Specifically,
we can calculate the non-interacting response function from the Lehmann representation in eq. (\ref{harm_harm_dens_resp_non_int_as_sum}) as
\begin{widetext}
\begin{gather}
	\chi_{\sigma \sigma}^{\mathrm{non-int}} (x, x', \omega)
	=
		\delta_{\sigma \sigma'}
		\frac{m \omega_0}{\pi}
		e^{-(z^2 + z'^2)/2}
		\sum_{n = 0}^\infty
			\frac{1}{2^n n !}
			H_n \left( \frac{z}{\sqrt{2}} \right)
			H_n \left( \frac{z'}{\sqrt{2}} \right)
			\left(
				\frac{1}{\omega - n \omega_0 + i \delta}
				-
				\frac{1}{\omega + n \omega_0 + i \delta}
			\right)
	,
	\label{harm_harm_dens_resp_non_int_Hermite}
\end{gather}
from which the coefficients are immediately found to be
\begin{gather}
	c_{\sigma n 0} (\sigma', x', \omega)
	=
		\frac{\delta_{\sigma \sigma'}}{2}
		\frac{e^{-z'^2/2}}{2^n n! \sqrt{\pi}}
		H_n \left( \frac{z'}{\sqrt{2}} \right)
		\left(
			\frac{1}{\omega/\omega_0 - n + i \delta}
			-
			\frac{1}{\omega/\omega_0 + n + i \delta}
		\right)
	.
	\label{harm_harm_WDF_exact_coeff_non_int}
\end{gather}
\end{widetext}
The expressions for the Kohn-Sham response function $\chi_{\sigma \sigma'}^{\mathrm{KS}}$ are obtained
simply by replacing $\omega_0$ with $\alpha^2 \omega_0$ in the expressions provided here.

\renewcommand{\refname}{ref}

\end{document}